\documentclass[11pt,a4paper]{article}
\pdfoutput=1
\usepackage{jheppub}
\usepackage{tikz}
\usetikzlibrary{intersections, calc, arrows.meta}
\usepackage{subcaption}
\usepackage{bm}

\DeclareMathOperator{\Tr}{Tr}

\newcommand{\ri}{\mathrm{i}}
\renewcommand{\th}{\theta}

\newcommand{\cob}{\delta}

\newcommand{\hf}{\frac{1}{2}}
\newcommand{\qu}{\frac{1}{4}}

\newcommand{\del}{\partial}

\newcommand{\bra}{\langle}
\newcommand{\ket}{\rangle}

\newcommand{\bt}{\beta}
\newcommand{\ga}{\gamma}
\newcommand{\Ga}{\Gamma}

\newcommand{\om}{\omega}

\newcommand{\rt}[1]{\sqrt{#1}}
\newcommand{\cO}{\mathcal{O}}

\newcommand{\bbZ}{{\mathbb Z}}
\newcommand{\gs}{g_{\rm s}}
\newcommand{\conn}{{\rm c}}
\newcommand{\ksi}{k_{\rm s}}
\newcommand{\kco}{k_{\rm c}}
\newcommand{\tG}{\widetilde{G}}
\newcommand{\tK}{\widetilde{K}}
\newcommand{\zt}{z_\tau}
\newcommand{\ta}{\tilde{a}}
\DeclareMathOperator{\Erf}{Erf}

\begin{document}

\title{\mathversion{bold}Spectral form factor in the $\tau$-scaling limit}

\author[a]{Kazumi Okuyama}
\author[b]{and Kazuhiro Sakai}

\affiliation[a]{Department of Physics, Shinshu University,\\
3-1-1 Asahi, Matsumoto 390-8621, Japan}
\affiliation[b]{Institute of Physics, Meiji Gakuin University,\\
1518 Kamikurata-cho, Totsuka-ku, Yokohama 244-8539, Japan}

\emailAdd{kazumi@azusa.shinshu-u.ac.jp, kzhrsakai@gmail.com}

\abstract{
We study the spectral form factor (SFF) of general topological gravity
in the limit of large time and fixed temperature. It has been observed
recently that in this limit, called the tau-scaling limit,
the genus expansion of the SFF can be summed up
and the late-time behavior of the SFF such as the ramp-plateau
transition can be studied analytically.
In this paper we develop a technique
for the systematic computation of the higher order corrections
to the SFF in the strict tau-scaling limit.
We obtain the first five corrections in a closed form
for the general background of topological gravity.
As concrete examples, we present the results for the Airy case and
Jackiw-Teitelboim gravity.
We find that the above higher order corrections
are the Fourier transforms of the corrections
to the sine-kernel approximation of the Christoffel-Darboux kernel
in the dual double-scaled matrix integral,
which naturally explains their structure.
Along the way we also develop a technique for
the systematic computation of the corrections
to the sine-kernel formula, which have not been
fully explored in the literature before.}

\maketitle

\section{Introduction}
Spectral form factor (SFF) is a useful measure of the level statistics
of quantum chaotic systems \cite{sff}
and it is widely studied in many areas of physics. 
In the context of quantum gravity and holography, 
the SFF of the Sachdev-Ye-Kitaev (SYK) model
\cite{Sachdev1993,Kitaev1,Kitaev2}
was studied in \cite{Garcia-Garcia:2016mno,Cotler:2016fpe} as a useful 
diagnostics of the Maldacena's version of the information
problem \cite{Maldacena:2001kr}.
It is found that the SFF of the SYK model 
exhibits the behavior of
the ramp and plateau as a function of time, 
which is consistent with the
conjecture
that the level statistics
of quantum chaotic system is universally described by 
a random matrix model \cite{bohigas1984}.
As shown in \cite{Saad:2019lba}, Jackiw-Teitelboim (JT)
gravity \cite{Jackiw:1984je,Teitelboim:1983ux}, 
which is holographically dual to  the low energy sector
of the SYK model, is indeed described by a 
certain double-scaled random matrix model. 
In the bulk gravity picture, the ramp of the
SFF of JT gravity comes from the contribution
of a wormhole connecting two boundaries of spacetime \cite{Saad:2018bqo}.
In the matrix model picture, it is described by the connected part of
the correlator of two
macroscopic loop operators \cite{Banks:1989df}.
On the other hand, the plateau of the SFF is very mysterious from 
the viewpoint of bulk gravity and it was speculated 
that
the appearance of the plateau is related
to some non-perturbative effects in quantum gravity 
\cite{Cotler:2016fpe,Saad:2019lba}. 
In the matrix model picture, it is argued in
\cite{Cotler:2016fpe,Saad:2019lba} that 
the plateau can be explained by the Andreev--Altshuler instantons
\cite{andreev1995spectral}.
See also \cite{Okuyama:2018gfr} for an explanation
of the plateau by eigenvalue instantons.

In this paper, we will consider the connected part of the SFF
in 2d quantum gravity
\begin{equation}
\begin{aligned}
 \text{SFF}=\bra Z(\bt+\ri t)Z(\bt-\ri t)\ket_\conn,
\end{aligned} 
\label{eq:SFF-def}
\end{equation}
where $Z(\bt)=\Tr e^{-\bt H}$ and the expectation value is defined
by averaging over a random matrix $H$.
As shown in \cite{Brezin:1990rb,Douglas:1989ve,Gross:1989vs},
2d quantum gravity coupled to a conformal matter can be defined
by a certain double-scaling limit of a random matrix model.
More generally, one can introduce couplings
$\{t_k\}\ (k=0,1,2,\ldots)$
to the matrix model potential. Then the free energy of the model 
is interpreted as a generating function of the intersection numbers
on the moduli space of Riemann surfaces and the model
is called topological gravity \cite{Witten:1990hr}.
There is an underlying integrable structure
in this model and the free energy serves as a tau-function of 
the KdV hierarchy \cite{Witten:1990hr,Kontsevich:1992ti}.
It turns out that the matrix model of JT gravity
in \cite{Saad:2019lba} is a special case of topological gravity
where infinitely many couplings $\{t_k\}$ are turned on
in a specific way \cite{mulase,Dijkgraaf:2018vnm,Okuyama:2019xbv}
and it corresponds to the computation of
Weil-Petersson volumes \cite{mirzakhani2007simple,Eynard:2007fi}.

SFF exhibits the ramp-plateau transition
around the time scale $t\sim \hbar^{-1}$, called the Heisenberg 
time, where $\hbar$  is the genus-counting parameter
of the matrix model.
Recently, it is observed in 
\cite{Saad:2022kfe,Blommaert:2022lbh,Weber:2022sov} that 
one can focus on the ramp-plateau transition regime
by taking what is called the
``$\tau$-scaling limit''\footnote{
In \cite{Okuyama:2019xbv,Okuyama:2020ncd}, another scaling limit
where both $\bt$ and $t$ are of order $\hbar^{-1}$ was
considered. This limit was called the
``'t Hooft limit'' in \cite{Okuyama:2019xbv}.
}
\begin{equation}
\begin{aligned}
 t\to\infty,\quad \hbar\to 0,\quad \tau=t\hbar: \text{fixed},
\end{aligned} 
\label{eq:tau-scaling}
\end{equation}
with $\bt$ fixed in \eqref{eq:SFF-def}.
In this limit, SFF is expanded as
\begin{equation}
\begin{aligned}
 \text{SFF}=\sum_{g=0}^\infty \hbar^{2g-1}\text{SFF}_g(\tau,\bt).
\end{aligned} 
\label{eq:SFFg-def}
\end{equation}
Remarkably, it is found that
the leading term $\text{SFF}_0$
can be computed in a closed form by just summing over the
original genus expansion in the 
$\tau$-scaling limit and the resulting
$\text{SFF}_0$ approaches a constant as $\tau\to\infty$ 
\cite{Saad:2022kfe}.
This opens up an interesting avenue for a ``perturbative plateau.''

In this paper, we will develop a technique for 
the systematic computation of the higher order corrections
$\text{SFF}_g~(g\geq1)$.
By using our method, we obtain $\text{SFF}_g$
up to $g=5$ for arbitrary couplings $\{t_k\}$ of topological
gravity.
It turns out that $\text{SFF}_g$ has a structure
which is a natural generalization of $\text{SFF}_0$.
In \cite{Saad:2022kfe} it was shown that $\text{SFF}_0$ is essentially determined
by the Fourier transform of the universal part
of the two-body eigenvalue correlation, known as the sine kernel
formula \cite{Gaudin:1961,Dyson:1962es,Brezin:1993qg}.
We find that the higher order correction
$\text{SFF}_g$ is closely related to the
correction of the Christoffel-Darboux (CD) kernel to 
the naive sine kernel formula. 
Rather surprisingly, such corrections to
the sine kernel formula have
not been fully explored in the literature before,
as far as we know.\footnote{
In \cite{Brezin:1993qg}, general form of 
the large $N$ limit of the CD kernel
was studied from the asymptotic behavior of the orthogonal
polynomials associated to an arbitrary matrix model potential,
before taking the double-scaling limit. 
\label{footnote}
}
We will also develop a technique for the systematic computation of the
corrections of the CD kernel and confirm that
$\text{SFF}_g$ is correctly reproduced form the Fourier transform of
the CD kernel by including the corrections to the sine kernel
formula.

This paper is organized as follows.
In section~\ref{sec:review},
we review the known results about the SFF in the $\tau$-scaling limit.
Along the way we consider the $\tau$-scaling limit
based on the genus expansion of the SFF
and obtain the small $\tau$ expansion of $\text{SFF}_g$ for small $g$.
In section~\ref{sec:summary}, we summarize our results of
$\text{SFF}_g$ and the higher order corrections of the CD kernel.
We also explain their relations.
In section~\ref{sec:example},
based on our general results
we compute $\text{SFF}_0$ and $\text{SFF}_1$
for the Airy case and JT gravity, as concrete examples.
In section~\ref{sec:SFF},
we formulate a systematic method of computing $\text{SFF}_g$.
In section~\ref{sec:CD},
we explain how to compute the higher order corrections of
the CD kernel beyond the sine kernel approximation.
Finally we conclude in section~\ref{sec:conclusion}.
In Appendix~\ref{app:airy-kernel}, we compute the
corrections of the Airy kernel to the sine kernel formula.
In Appendix~\ref{app:a2},
we present an explicit form of the coefficient
that determines $\text{SFF}_2$.

\section{\mathversion{bold}$\tau$-scaling limit of SFF}
\label{sec:review}

In this section we will briefly review the known results about
the SFF in the $\tau$-scaling limit.
We will consider the $\tau$-scaling limit
for the general background of topological gravity
based on the known genus expansion of the SFF,
which enables us to compute the small $\tau$ expansion of $\text{SFF}_g$
for small $g$.

For the general background $\{t_k\}$ of
topological gravity, the connected two-boundary correlator
is expanded as \cite{Okuyama:2020ncd}
\begin{equation}
\begin{aligned}
 \bra Z(\bt_1)Z(\bt_2)\ket_\conn&=\frac{\rt{\bt_1\bt_2}}{2\pi}e^{-(\bt_1+\bt_2)E_0}\Biggl[
\frac{1}{\bt_1+\bt_2}\\
&\qquad+\left(\frac{\bt_1^2+\bt_1\bt_2+\bt_2^2}{24s^2}
+\frac{2(\bt_1+\bt_2)I_2+I_3}{24s^3}+\frac{I_2^2}{12s^4}\right)\gs^2+\cO(\gs^4)\Biggr],
\end{aligned} 
\label{eq:ZZ-expand}
\end{equation}
where $\gs$ is the genus-counting parameter related to $\hbar$
as\footnote{See \cite{Okuyama:2019xbv} for more about 
our definition of
$\hbar$ and its relation to the parameters $S_0,\ga$ in JT gravity.
}
\begin{equation}
\begin{aligned}
 \gs=\rt{2}\hbar.
\end{aligned} 
\end{equation}
$I_k$ denote the Itzykson-Zuber variables defined by
\cite{Itzykson:1992ya}
\begin{equation}
\begin{aligned}
 I_k=\sum_{n=0}^\infty \frac{t_{k+n}}{n!}(-E_0)^n,
\quad k\in\bbZ_{\ge 0}
\end{aligned} 
\end{equation}
and $s$ in \eqref{eq:ZZ-expand} is given by
\begin{equation}
\begin{aligned}
 s=1-I_1.
\end{aligned} 
\end{equation}
The threshold energy $E_0$ is determined by the (genus-zero) string equation
\begin{equation}
\begin{aligned}
 I_0+E_0=0.
\end{aligned} 
\end{equation}
In the $\tau$-scaling limit \eqref{eq:tau-scaling},
after setting $\bt_{1,2}=\bt\pm\ri \tau/\hbar$ in the two-boundary correlator
\eqref{eq:ZZ-expand} and expanding it in $\hbar$,
we find the small $\tau$ expansion of $\text{SFF}_{0}$ and $\text{SFF}_{1}$
\begin{equation}
\begin{aligned}
 \text{SFF}_{0}&=\frac{e^{-2\bt E_0}}{2\pi}\left[
\frac{\tau}{2 \bt}-\frac{\tau^3}{12 s^2}+\cdots\right],\\
\text{SFF}_{1}&=\frac{e^{-2\bt E_0}}{2\pi}\left[
\frac{\bt}{4 \tau}+ \left(\frac{5 \bt^2}{24 s^2}+\frac{4\bt I_2+I_3}{12 s^3}
+\frac{I_2^2}{6 s^4}\right)\tau+\cdots\right].
\end{aligned} 
\label{eq:SFF-small-tau}
\end{equation}

In general, the two-boundary correlator is written as
\begin{equation}
\begin{aligned}
 \bra Z(\bt_1)Z(\bt_2)\ket_\conn&=\bra Z(\bt_1+\bt_2)\ket-
\int dE_1dE_2 e^{-\bt_1E_1-\bt_2E_2}K(E_1,E_2)^2,
\end{aligned} 
\label{eq:ZZ-b1b2}
\end{equation}
where $K(E_1,E_2)$ is the Christoffel-Darboux (CD) kernel.
Thus, the SFF is written as
\begin{equation}
\begin{aligned}
 \text{SFF}&=\int\frac{dE}{2\pi}e^{-2\bt E}\rho(E)-
\int dE_1dE_2 e^{-\bt(E_1+E_2)-\ri\tau \frac{E_1-E_2}{\hbar}}K(E_1,E_2)^2.
\end{aligned} 
\label{eq:SFF-general}
\end{equation}
In the $\tau$-scaling limit \eqref{eq:tau-scaling}, 
the above integral is dominated by the region
\begin{equation}
\begin{aligned}
 E_1-E_2\sim \cO(\hbar),
\end{aligned} 
\end{equation}
with finite $E_1+E_2$. In this regime, the CD kernel is approximated by the universal
two-body correlation of eigenvalues, known as the sine kernel
\cite{Gaudin:1961,Dyson:1962es,Brezin:1993qg}
\begin{equation}
\begin{aligned}
 K_{\text{sin}}(E_1,E_2)
=\frac{\sin\bigl[\frac{1}{2}\rho_0(E)\om\bigr]}{\pi\hbar\om},
\end{aligned} 
\label{eq:sine-kernel}
\end{equation}
where we defined
\begin{equation}
\begin{aligned}
 E_1=E+\hf\hbar\om,\quad E_2=E-\hf\hbar\om.
\end{aligned} 
\label{eq:E1E2}
\end{equation}
$\rho_0(E)$ is the genus-zero part of the
eigenvalue density
\begin{equation}
\begin{aligned}
\rho(E)=\sum_{g=0}^\infty \hbar^{2g-1}\rho_g(E)
\end{aligned} 
\end{equation}
which is related to the one-point function by
\begin{equation}
\begin{aligned}
 \bra Z(\bt)\ket=\int_{E_0}^\infty \frac{dE}{2\pi}e^{-\bt E}\rho(E)
=\sum_{g=0}^\infty\hbar^{2g-1}\bra Z(\bt)\ket_g.
\end{aligned} 
\label{eq:Zrhorel}
\end{equation}
For the general background $\{t_k\}$, $\rho_0(E)$ is given by
\begin{equation}
\begin{aligned}
  \rho_0(E)=\sum_{k=1}^\infty 
\frac{(-1)^k (I_k-\cob_{k,1})\Ga(1/2)}{\Ga(k+1/2)}(E-E_0)^{k-\hf}.
\end{aligned} 
\label{eq:rho0}
\end{equation}

In \cite{Saad:2022kfe,Blommaert:2022lbh,Weber:2022sov}, it is argued that
the above small $\tau$ expansion of 
$\text{SFF}_0$ \eqref{eq:SFF-small-tau}
can be resummed and 
$\text{SFF}_0$ is obtained by replacing the CD kernel by the sine kernel
in \eqref{eq:SFF-general}.
Then, by the change of integration variables in \eqref{eq:E1E2},
the leading term of the SFF becomes
\begin{equation}
\begin{aligned}
 \text{SFF}_0&=\int_{E_0}^\infty \frac{dE}{2\pi}e^{-2\bt E}\rho_0(E)
-\int_{E_0}^\infty dEe^{-2\bt E}\int_{-\infty}^\infty d\om e^{-\ri\om\tau}
\frac{\sin^2\bigl(\hf\rho_0(E)\om\bigr)}{\pi^2\om^2}.
\end{aligned} 
\label{eq:SFF0-int}
\end{equation}
The $\om$-integral is evaluated as \cite{brezin-hikami}
\begin{equation}
\begin{aligned}
 \int_{-\infty}^\infty d\om e^{-\ri\om\tau}
\frac{\sin^2\bigl[\hf\rho_0(E)\om\bigr]}{\pi^2\om^2}=
\frac{1}{2\pi}\bigl(\rho_0(E)-\tau\bigr)\th\bigl(\rho_0(E)-\tau\bigr),
\end{aligned} 
\end{equation}
where $\th(x)$ is the step function
\begin{equation}
\th(x)=\left\{
\begin{aligned}
 &1,\quad &(x>0),\\
&0,\quad&(x<0).
\end{aligned}\right. 
\end{equation}
Assuming that $\rho_0(E)$ is a monotonically increasing function
of $E$, there is a unique solution $E_\tau$ to the equation
\begin{equation}
\begin{aligned}
 \rho_0(E_\tau)=\tau.
\end{aligned} 
\label{eq:Etau-def}
\end{equation}
In terms of $E_\tau$, $\text{SFF}_0$ in \eqref{eq:SFF0-int} is written as
\begin{equation}
\begin{aligned}
 \text{SFF}_0&=\int_{E_0}^\infty \frac{dE}{2\pi}e^{-2\bt E}\rho_0(E)
-\int_{E_\tau}^\infty \frac{dE}{2\pi}e^{-2\bt E}\bigl(\rho_0(E)-\tau\bigr)\\
&=\frac{\tau}{4\pi\bt}e^{-2\bt E_\tau}
+\int_{E_0}^{E_\tau} \frac{dE}{2\pi}e^{-2\bt E}\rho_0(E).
\end{aligned} 
\label{eq:SFF0}
\end{equation}
The first term corresponds to the ramp and the
second term approaches a constant $\bra Z(2\bt)\ket_0$
in the $\tau\to\infty$ limit, corresponding to the plateau. 

One can check that \eqref{eq:SFF0} reproduces 
the small $\tau$ expansion of $\text{SFF}_0$ in \eqref{eq:SFF-small-tau}.
To see this, we first notice that $E_\tau$ has the following small
$\tau$ expansion\footnote{A fully explicit expression of
this expansion is available
($E_\tau$ is given by $E(\lambda)$ in
\eqref{eq:Eofla} with $\lambda=\ri\tau$).}
\begin{equation}
\begin{aligned}
 E_\tau-E_0=\frac{1}{4s^2}\tau^2-\frac{I_2}{12s^5}\tau^4
+\left(\frac{7I_2^2}{144s^8}+\frac{I_3}{120s^7}\right)\tau^6+\cO(\tau^8).
\end{aligned} 
\end{equation}
This can be obtained by inverting the relation
\eqref{eq:Etau-def} using the expression of $\rho_0(E)$
in \eqref{eq:rho0}.
Plugging this expansion of $E_\tau$ into
\eqref{eq:SFF0}, one can show that the 
small $\tau$ expansion of $\text{SFF}_0$ in \eqref{eq:SFF-small-tau}
is indeed reproduced from \eqref{eq:SFF0}.

In the rest of this paper, we will consider the higher order corrections 
$\text{SFF}_g~(g\geq1)$ to the spectral form factor.
In the next section we will first summarize the result of $\text{SFF}_g$.
The details of the computation will be postponed to section~\ref{sec:SFF}.

\section{Summary of results}\label{sec:summary}

In this section we will summarize our results of
$\text{SFF}_g$ and discuss their relation to
the corrections of the CD kernel.

\subsection{$\text{SFF}_g$}\label{sec:SFFg}

It turns out that the expression of $\text{SFF}_0$ in \eqref{eq:SFF0}
has a natural generalization to the higher order correction $\text{SFF}_g$.
We find that $\text{SFF}_g$ has the structure
\begin{equation}
\begin{aligned}
 \text{SFF}_g=\frac{1}{2\pi}f_g(\tau,\bt)e^{-2\bt E_\tau}+
\int_{E_0}^{E_\tau}\frac{dE}{2\pi}e^{-2\bt E}\rho_g(E).
\end{aligned} 
\label{eq:SFFg-form}
\end{equation} 
By the relation \eqref{eq:Zrhorel}
we can easily find the first few terms of $\rho_g(E)$
from the result of $\bra Z(\bt)\ket_g$ in \cite{Okuyama:2019xbv}.
For instance,
\begin{equation}
\begin{aligned}
 \rho_1(E)&=\frac{1}{16sz^5}-\frac{I_2}{24s^2z^3},\\
\rho_2(E)&=
-\frac{105}{1024 s^3z^{11}}+\frac{203 I_2}{1536 s^4z^9}\\
&\quad+\frac{1}{z^7}\left(
-\frac{7 I_2^2}{64 s^5}-\frac{29 I_3}{768 s^4}\right)
+\frac{1}{z^5}\left(\frac{7 I_2^3}{96 s^6}+\frac{29 I_2 I_3}{480 s^5}+\frac{I_4}{128 s^4}\right)\\
&\quad+\frac{1}{z^3}\left(-\frac{7 I_2^4}{144 s^7}-\frac{5 I_2^2 I_3}{72 s^6}-\frac{11 I_2 I_4}{720
   s^5}-\frac{29 I_3^2}{2880 s^5}-\frac{I_5}{576 s^4}\right),
\end{aligned} 
\label{eq:rho1and2}
\end{equation}
where
\begin{equation}
\begin{aligned}
 z=\rt{E-E_0}.
\end{aligned} 
\end{equation}
Note that the second term of \eqref{eq:SFFg-form} is a formal expression since
the integral has a divergence coming from $E=E_0$.
This integral should be understood by a certain analytic continuation.
For instance, the integral involving $z^{-a}$ term is defined as
\begin{equation}
\begin{aligned}
 \int_{E_0}^{E_\tau} dEe^{-2\bt E}z^{-a}&=
\int_{E_0}^{\infty} dEe^{-2\bt E}z^{-a}+\int_{\infty}^{E_\tau} dEe^{-2\bt E}z^{-a}\\
&=e^{-2\bt E_0}\Ga(1-a/2)(2\bt)^{a/2-1}+\int_{\infty}^{E_\tau} dEe^{-2\bt E}z^{-a}.
\end{aligned} 
\end{equation}
Using this prescription, one can show that the second term of 
\eqref{eq:SFFg-form} is written as
\begin{equation}
\begin{aligned}
 \int_{E_0}^{E_\tau}\frac{dE}{2\pi}e^{-2\bt E}\rho_g(E)&=\bra Z(2\bt)\ket_g
\text{Erf}(z_\tau\rt{2\bt})+\frac{1}{2\pi}d_g(z_\tau,\bt)e^{-2\bt E_\tau},
\end{aligned}
\label{eq:ZErf-cg} 
\end{equation}
where $\text{Erf}(z)$ denotes the error function and $z_\tau$ is defined by
\begin{equation}
\begin{aligned}
 z_\tau=\rt{E_\tau-E_0}.
\end{aligned} 
\end{equation}
It turns out that $d_g(z_\tau,\bt)$ in \eqref{eq:ZErf-cg} 
is a polynomial in $1/z_\tau$.
For instance, $d_1(z_\tau,\bt)$ is given by
\begin{equation}
\begin{aligned}
 d_1(z_\tau,\bt)&=\frac{\bt}{6sz_\tau}-\frac{1}{24sz_\tau^3}+\frac{I_2}{12s^2z_\tau}.
\end{aligned} 
\label{eq:nonErf1}
\end{equation}

Next, let us consider the first term of \eqref{eq:SFFg-form}.
We find that $f_g(\tau,\bt)e^{-2\bt E_\tau}$ can be rewritten as
a sum of $\tau$-derivatives
\begin{equation}
\begin{aligned}
 f_g(\tau,\bt)e^{-2\bt E_\tau}=\sum_{n=0}^{3g-2} \del_\tau^n
\bigl[h_{g,n}(\tau)e(1)e^{-2\bt E_\tau}\bigr].
\end{aligned} 
\label{eq:fg-str}
\end{equation}
Here we introduced the notation
\begin{equation}
\begin{aligned}
 e(j)=\del_\tau^jE_\tau.
\end{aligned} 
\label{eq:ej-def}
\end{equation}
The important point is that
$h_{g,n}(\tau)$ is independent of $\bt$; the $\bt$-dependence of
$f_g(\tau,\bt)$ arises solely from the $\tau$-derivative in  
\eqref{eq:fg-str}. We have computed $h_{g,n}$ up to $g=5$.
For $g=1$ we find
 \begin{equation}
\begin{aligned}
 h_{1,1}=-s(2),\quad
h_{1,0}=\frac{1}{16z_\tau^4},
\end{aligned} 
\label{eq:h1n}
\end{equation}
where $s(j)$ denotes
\begin{equation}
\begin{aligned}
 s(j)=\frac{\rho_0^{(j)}(E_\tau)}{(j+1)!2^{j}}
\end{aligned} 
\end{equation}
with $\rho_0^{(j)}(E)=\del_E^j\rho_0(E)$. For $g=2$ we find
\begin{equation}
\begin{aligned}
  h_{2,4}&=-\hf s(2)^2,\\
h_{2,3}&=s(4)+\frac{1}{16z_\tau^4}s(2),\\
h_{2,2}&=-\frac{3}{256z_\tau^8}+\rho_1(E_\tau) s(2),\\
h_{2,1}&=\frac{7 I_2}{768 s^2 z_\tau^7}-\frac{73}{1536 s z_\tau^9},\\
h_{2,0}&=-\frac{I_3}{96 s^3 z_\tau^6}-\frac{7}{128 s^2 z_\tau^{10}}-\frac{25}{2}
\rho_1(E_\tau)^2.
\end{aligned} 
\label{eq:h2n}
\end{equation}
From the definition of $E_\tau$ in \eqref{eq:Etau-def}, one can show that
$\rho_0^{(j)}(E_\tau)$ can be written as a combination of $e(j)$. For instance,
\begin{equation}
\begin{aligned}
\rho_0^{(2)}(E_\tau)&=-\frac{e(2)}{e(1)^3},\\
\rho_0^{(4)}(E_\tau)&=
-\frac{e(4)}{e(1)^5}-\frac{15 e(2)^3}{e(1)^7}+\frac{10 e(3) e(2)}{e(1)^6}.
\end{aligned} 
\end{equation}
For general $g$, we find that $h_{g,n}$ with $n=3g-2$ and $n=3g-3$ are given by
\begin{equation}
\begin{aligned}
 h_{g,3g-2}=-\frac{s(2)^g}{g!},\quad
h_{g,3g-3}=\frac{1}{16z_\tau^4}\frac{s(2)^{g-1}}{(g-1)!}
+s(4)\frac{s(2)^{g-2}}{(g-2)!}.
\end{aligned} 
\end{equation} 
In subsection~\ref{sec:sine-correctio}, we will see that this structure
can be naturally understood from the correction
of the CD kernel to the sine kernel formula.

Let us take a closer look at the behavior of $\text{SFF}_g$ in \eqref{eq:SFFg-form}
as a function of $\tau$. At late times, the first term of 
\eqref{eq:SFFg-form} vanishes exponentially and the second term approaches
a constant
\begin{equation}
\begin{aligned}
 \lim_{\tau\to\infty}\text{SFF}_g=\bra Z(2\bt)\ket_g.
\end{aligned} 
\end{equation}
This gives the higher genus correction to the value of plateau.
On the other hand, at early times
$\text{SFF}_g$ diverges as
\begin{equation}
\begin{aligned}
\lim_{\tau\to0}\text{SFF}_g \sim e^{-2\bt E_0}\left(\frac{\bt}{\tau}\right)^{2g-1}.
\end{aligned} 
\end{equation} 
This is just an artifact of the $\tau$-scaling limit.
This early time divergence can be traced back to the expansion of the
original genus-zero term
\begin{equation}
\begin{aligned}
 \frac{\rt{\bt_1\bt_2}}{2\pi(\bt_1+\bt_2)}&=\frac{\rt{\tau^2+\bt^2\hbar^2}}{4\pi\bt\hbar}
=\frac{1}{4\pi}\sum_{g=0}^\infty \frac{(-1)^{g-1}(2g-3)!!}{g!2^g}
\left(\frac{\bt\hbar}{\tau}\right)^{2g-1}.
\end{aligned} 
\label{eq:early-time}
\end{equation}
Before expanding in $\hbar$, the original 
expression $\rt{\tau^2+\bt^2\hbar^2}$ is regular at $\tau=0$,
but after expanding it by $\hbar$ there appears an apparent divergence at $\tau=0$.
We have checked that the coefficient of $\cO(\tau^{1-2g})$ term of
$\text{SFF}_g$ in the small $\tau$ expansion is indeed given by 
\eqref{eq:early-time}.

\subsection{Relation to the corrections of CD kernel}\label{sec:sine-correctio}
From the general relation between the SFF and the CD kernel 
$K(E_1,E_2)$ in \eqref{eq:SFF-general},
the higher order correction $\text{SFF}_g$ is closely related to the
correction of the CD kernel to the naive sine kernel formula 
\eqref{eq:sine-kernel}. The details of the computation of CD kernel will be presented in section~\ref{sec:CD}.
It turns out that the $\tau$-derivative structure in \eqref{eq:fg-str}
naturally appears from the Fourier transform of 
the CD kernel squared
\begin{equation}
\begin{aligned}
 \text{SFF}=\int\frac{dE}{2\pi}e^{-2\bt E}
\rho(E)-\hbar\int dE e^{-2\bt E}\int d\om e^{-\ri\om\tau}
K\left(E+\hf\hbar\om,E-\hf\hbar\om\right)^2.
\end{aligned} 
\label{eq:SFF-gen2}
\end{equation}
We find that the corrections of the CD kernel to the sine kernel formula
is organized as
\begin{equation}
\begin{aligned}
K\left(E+\hf\hbar\om,E-\hf\hbar\om\right)&=\Biggl[\frac{2}{\hbar\om}
+\sum_{g=1}^\infty\hbar^{2g-1}\ksi^{(g)}(E,\om)\Biggr]\frac{\sin\phi}{2\pi}\\
&+\sum_{g=1}^\infty\hbar^{2g-1}\Bigl[\rho_g(E)+\kco^{(g)}(E,\om)\Bigr]
\frac{\cos\phi}{2\pi},
\end{aligned} 
\label{eq:CD-sincos}
\end{equation}
where $\phi$ is given by
\begin{equation}
\begin{aligned}
 \phi=\frac{1}{2\hbar}\int_{E-\hf\hbar\om}^{E+\hf\hbar\om}dE\rho_0(E)
=\hf\sum_{j=0}^\infty \hbar^{2j}\om^{2j+1}
\frac{\rho_0^{(2j)}(E)}{(2j+1)!2^{2j}}.
\end{aligned} 
\label{eq:phi}
\end{equation}
Note that the diagonal part of the CD kernel is equal to the eigenvalue density
\begin{equation}
\begin{aligned}
 K(E,E)=\frac{1}{2\pi}\rho(E)=\frac{1}{2\pi}\sum_{g=0}^\infty\hbar^{2g-1}\rho_g(E).
\end{aligned} 
\label{eq:K-diag}
\end{equation}
We find that $\ksi^{(g)}$ and $\kco^{(g)}$ vanishes as $\om\to0$ and 
hence our expansion \eqref{eq:CD-sincos} is consistent with the diagonal 
part \eqref{eq:K-diag}.
The explicit forms of $\ksi^{(g)}$ and $\kco^{(g)}$ for $g=1,2$ are given by 
\begin{equation}
\begin{aligned}
 \ksi^{(1)}&=\frac{\om}{16z^4},\\
\kco^{(1)}&=0,\\
\ksi^{(2)}&=\frac{11\om^3}{1024z^8}+\left(-\frac{7}{128 s^2 z^{10}}-\frac{I_3}{96
   s^3 z^6}-\frac{49}{4}\rho_1(E)^2\right)\om,\\
\kco^{(2)}&=\left(\frac{35}{768sz^9}-\frac{I_2}{128s^2z^7}\right)\om^2.
\end{aligned} 
\label{eq:kskc-g12}
\end{equation}
Expanding $\sin\phi$ and $\cos\phi$ further in $\hbar$, we find
the $\hbar$ expansion of the CD kernel
in \eqref{eq:CD-sincos}
\begin{equation}
\begin{aligned}
 K\left(E+\hf\hbar\om,E-\hf\hbar\om\right)=\sum_{g=0}^\infty \hbar^{2g-1}K_g(E,\om).
\end{aligned} 
\end{equation}
One can see that the leading term is the sine kernel
\begin{equation}
\begin{aligned}
 K_0(E,\om)=\frac{\sin\bigl[\hf\rho_0(E)\om\bigr]}{\pi\om},
\end{aligned} 
\end{equation}
and the higher order correction $K_g(E,\om)$ can be obtained from
\eqref{eq:CD-sincos}.

From the general form of the CD kernel \eqref{eq:CD-sincos}, one can see
that the $\tau$-derivative structure of \eqref{eq:fg-str}
can be understood from the Fourier transform of \eqref{eq:CD-sincos}.
Let us consider a contribution of the term
\begin{equation}
\begin{aligned}
K\left(E+\hf\hbar\om,E-\hf\hbar\om\right)^2\supset \om^n e^{\ri\rho_0(E)\om},\quad(n\geq0).
\end{aligned} 
\end{equation}
From the Fourier transformation of this term, we find
\begin{equation}
\begin{aligned}
 \int_{-\infty}^\infty \frac{d\om}{2\pi} e^{-\ri\om\tau}\om^n e^{\ri\rho_0(E)\om}&=(\ri\del_\tau)^n
\int_{-\infty}^\infty \frac{d\om}{2\pi} e^{-\ri\om\tau+\ri\rho_0(E)\om}\\
&=(\ri\del_\tau)^n\cob\bigl(\rho_0(E)-\tau\bigr)\\
&=(\ri\del_\tau)^n\bigl[e(1)\cob(E-E_\tau)\bigr],
\end{aligned} 
\end{equation}
where we used the relation
\begin{equation}
\begin{aligned}
 \rho^{(1)}(E_\tau)=\frac{1}{e(1)}.
\end{aligned} 
\end{equation}
After integrating over $E$ in \eqref{eq:SFF-gen2}, we find the
$\tau$-derivative structure of $f_g(\tau,\bt)e^{-2\bt E_\tau}$ in
\eqref{eq:fg-str}.
We have checked that $h_{1,n}$ in \eqref{eq:h1n} and $h_{2,n}$ in \eqref{eq:h2n}
are correctly reproduced from the Fourier transform of
the square of the CD kernel in \eqref{eq:CD-sincos}.
In particular, the appearance of $s(2j)$ can be naturally understood from the
expansion of $\phi$ in \eqref{eq:phi}.

The second term of \eqref{eq:SFFg-form} arises from the
cross-term of $\sin\phi$ and $\cos\phi$ in the CD kernel squared
\begin{equation}
\begin{aligned}
 K\left(E+\hf\hbar\om,E-\hf\hbar\om\right)^2\supset
\frac{\sin\bigl[\rho_0(E)\om\bigr]}{2\pi^2\hbar\om}\sum_{g=1}^\infty
\hbar^{2g-1}\rho_g(E).
\end{aligned} 
\end{equation}
Using the relation
\begin{equation}
\begin{aligned}
 \int_{-\infty}^\infty d\om e^{-\ri\om\tau}\frac{\sin\bigl[\rho_0(E)\om\bigr]}{2\pi^2\om}
=\frac{1}{2\pi}\th\bigl(\rho_0(E)-\tau\bigr),
\end{aligned} 
\end{equation}
we find that the second term of \eqref{eq:SFFg-form}
is reproduced from \eqref{eq:SFF-gen2}
\begin{equation}
\begin{aligned}
 \text{SFF}_g&\supset \int_{E_0}^\infty\frac{dE}{2\pi}e^{-2\bt E}\rho_g(E)
-\int_{E_0}^\infty\frac{dE}{2\pi}e^{-2\bt E}\rho_g(E)\th\bigl(\rho_0(E)-\tau\bigr)\\
&=\int_{E_0}^\infty\frac{dE}{2\pi}e^{-2\bt E}\rho_g(E)
-\int_{E_\tau}^\infty\frac{dE}{2\pi}e^{-2\bt E}\rho_g(E)\\
&=\int_{E_0}^{E_\tau}\frac{dE}{2\pi}e^{-2\bt E}\rho_g(E).
\end{aligned} 
\end{equation}
To summarize, the structure of $\text{SFF}_g$ in \eqref{eq:SFFg-form}
and \eqref{eq:fg-str} can be naturally understood from the Fourier transform
of the CD kernel in \eqref{eq:CD-sincos}.
To our knowledge, corrections to the sine kernel formula
are not known in the literature before
and our \eqref{eq:CD-sincos} with \eqref{eq:kskc-g12}
is a new result.\footnote{
See also a comment in footnote \ref{footnote}.
}

\section{Examples of Airy case and JT gravity}\label{sec:example}

In this section, as concrete examples we compute $\text{SFF}_0$ and
$\text{SFF}_1$ for the Airy case and JT gravity
using our general result in the previous section.  

\subsection{Airy case}
Let us first consider the Airy case.
For the Airy case, all couplings $t_k$ are set to zero.
Then one can show that $E_0=0$ and $I_k=0~(k\geq1)$,
and the genus-zero eigenvalue density \eqref{eq:rho0} becomes
\begin{equation}
\begin{aligned}
 \rho_0(E)=2\rt{E}.
\end{aligned} 
\end{equation} 
From the definition of $E_\tau$ in \eqref{eq:Etau-def},
we find
\begin{equation}
\begin{aligned}
 E_\tau=\frac{\tau^2}{4},\quad z_\tau=\rt{E_\tau}=\frac{\tau}{2}.
\end{aligned} 
\end{equation}
From the general formula in the previous sections,
we find that the $\text{SFF}_0$ and
$\text{SFF}_1$ for the Airy case are given by
\begin{equation}
\begin{aligned}
 \text{SFF}_0&=
\frac{1}{2\rt{\pi(2\bt)^3}}\text{Erf}\left(\tau\rt{\frac{\bt}{2}}\right),\\
\text{SFF}_1&=\frac{\bt}{8\pi\tau}e^{-\hf\bt\tau^2}+
\frac{\bt^{3/2}}{6\rt{2\pi}}\text{Erf}\left(\tau\rt{\frac{\bt}{2}}\right).
\end{aligned} 
\label{eq:SFF-Airy}
\end{equation}
We can compare this with the exact result of two-boundary
correlator for the Airy case \cite{okounkov2002generating}
\begin{equation}
\begin{aligned}
 \bra Z(\bt_1)Z(\bt_2)\ket_\conn=\frac{e^{\frac{\hbar^2}{12}(\bt_1+\bt_2)^3}}{2\hbar\rt{\pi(\bt_1+\bt_2)^3}}\text{Erf}\left(\frac{\hbar}{2}\rt{\bt_1\bt_2(\bt_1+\bt_2)}\right).
\end{aligned} 
\label{eq:Airy-exact}
\end{equation}
One can check that the $\text{SFF}_{0,1}$
in \eqref{eq:SFF-Airy} are indeed reproduced from the
$\tau$-scaling limit of the exact result \eqref{eq:Airy-exact}.
See also Appendix \ref{app:airy-kernel} for the corrections of 
the CD kernel in the Airy case.

\subsection{JT gravity}
\begin{figure}[t]
\centering
\subcaptionbox{$\text{SFF}_0$\label{sfig:SFF0}}{\includegraphics
[width=0.45\linewidth]{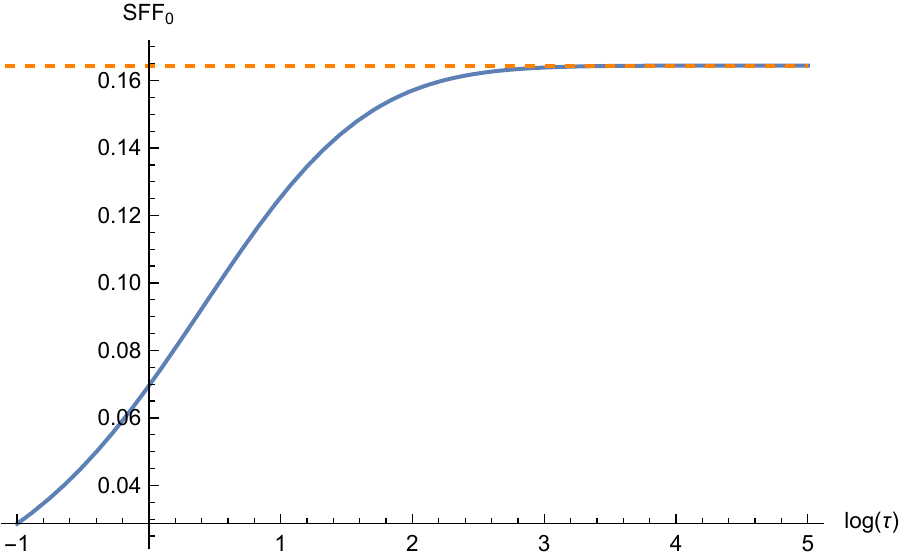}}
\hskip5mm
\subcaptionbox{$\text{SFF}_1$\label{sfig:SFF1}}{\includegraphics
[width=0.45\linewidth]{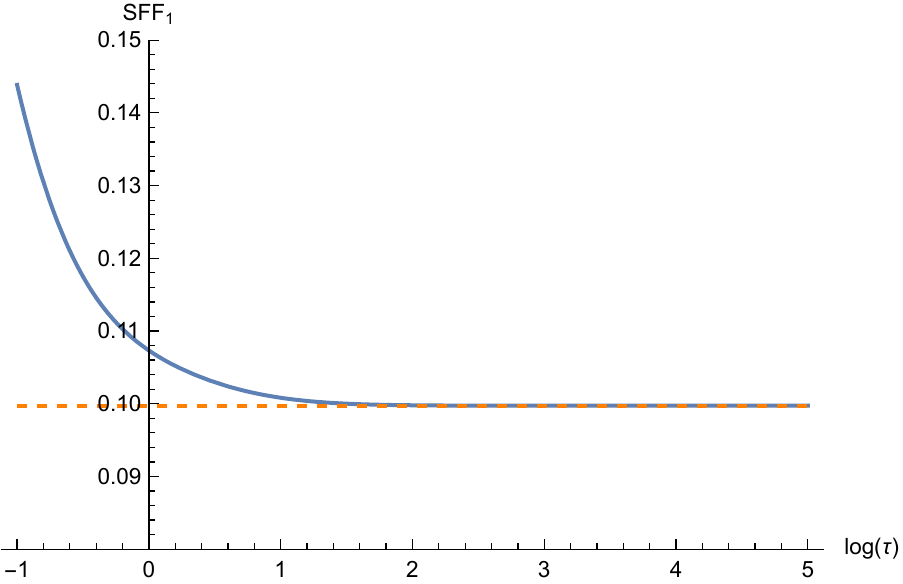}}
  \caption{
Plot 
of \subref{sfig:SFF0} $\text{SFF}_0$ and \subref{sfig:SFF1} $\text{SFF}_1$
for JT gravity. The horizontal axis is $\log\tau$.
We set $\bt=1$ in this figure. $\text{SFF}_g$ approaches $\bra Z(2\bt)\ket_g$
as $\tau\to\infty$ (shown by orange dashed lines).
}
  \label{fig:JT-SFF}
\end{figure}

Next consider the SFF of JT gravity.
For JT gravity, the couplings $t_k$ are given by
\cite{mulase,Dijkgraaf:2018vnm,Okuyama:2019xbv}
\begin{equation}
\begin{aligned}
 t_0=t_1=0,\qquad t_k=\frac{(-1)^k}{(k-1)!},~~(k\geq2).
\end{aligned} 
\end{equation}
In this case, $E_0=0$ and the genus-zero eigenvalue density \eqref{eq:rho0}
becomes
\begin{equation}
\begin{aligned}
 \rho_0(E)=\sinh(2\rt{E}).
\end{aligned} 
\end{equation}
From \eqref{eq:Etau-def}, $E_\tau$ and $z_\tau$ are given by
\begin{equation}
\begin{aligned}
 E_\tau=\qu\text{arcsinh}(\tau)^2,\quad z=\rt{E_\tau}=\hf\text{arcsinh}(\tau).
\end{aligned}
\label{eq:Etau-JT} 
\end{equation}
From the general formula \eqref{eq:SFF0},
the leading term $\text{SFF}_0$ is evaluated as \cite{Saad:2022kfe}
\begin{equation}
\begin{aligned}
 \text{SFF}_0
=\hf\bra Z(2\bt)\ket_{g=0}\left[
\text{Erf}\left(\frac{\bt \text{arcsinh}(\tau)+1}{\rt{2\bt}}\right)+
\text{Erf}\left(\frac{\bt \text{arcsinh}(\tau)-1}{\rt{2\bt}}\right)\right],
\end{aligned} 
\end{equation}
where the genus-zero one-point function is given by
\begin{equation}
\begin{aligned}
 \bra Z(\bt)\ket_{g=0}=\frac{e^{\frac{1}{\bt}}}{2\rt{\pi\bt^3}}.
\end{aligned} 
\end{equation}
The next order correction $\text{SFF}_1$ for JT gravity can be found from
the general result in the previous section. After some algebra, we find
\begin{equation}
\begin{aligned}
 \text{SFF}_1
&=\frac{1}{24\pi}e^{-\hf\bt\text{arcsinh}(\tau)^2}
\Biggl[\frac{\tau}{1+\tau^2}\left(\bt+\frac{1}{\text{arcsinh}(\tau)^2}\right)
+\frac{4}{\text{arcsinh}(\tau)^3}\left(\frac{1}{\rt{1+\tau^2}}-1\right)\\
&\quad+\frac{1}{\text{arcsinh}(\tau)}\left(2-\frac{1}{(1+\tau^2)^{3/2}}
+\bt\left(4-\frac{1}{\rt{1+\tau^2}}\right)\right)\Biggr]\\
&\qquad +\bra Z(2\bt)\ket_{g=1}\text{Erf}\left(\rt{\frac{\bt}{2}}
\text{arcsinh}(\tau)\right),
\end{aligned} 
\label{eq:SFF1-JT}
\end{equation}
where the genus-one one-point function is given by
\begin{equation}
\begin{aligned}
 \bra Z(\bt)\ket_{g=1}=\frac{(1+\bt)\rt{\bt}}{24\rt{\pi}}.
\end{aligned} 
\end{equation}
In Figure \ref{fig:JT-SFF}, we show the plot of 
$\text{SFF}_0$ and $\text{SFF}_1$ of JT gravity.
From Figure \ref{sfig:SFF0} we can see that
$\text{SFF}_0$ exhibits the behavior of the ramp and plateau.
From Figure \ref{sfig:SFF1} we see that $\text{SFF}_1$ approaches a constant at late times while it
blows up as $\tau\to0$. As we argued in the previous section, this diverging behavior
of $\text{SFF}_1$ at $\tau=0$ is an artifact of the $\tau$-scaling limit.
In fact, the small $\tau$ expansion of $\text{SFF}_1$ in \eqref{eq:SFF1-JT}
reads
\begin{equation}
\begin{aligned}
 \text{SFF}_1=\frac{1}{2\pi}\Biggl[\frac{\bt}{4\tau}+\frac{3+8\bt+5\bt^2}{24}\tau+\cO(\tau^3)\Biggr].
\end{aligned} 
\label{eq:SFF1-smalltau}
\end{equation}
One can see that the negative power term of $\tau$ agrees with the
$\cO(\hbar)$ term of \eqref{eq:early-time}, as expected. One can also check that
this expansion \eqref{eq:SFF1-smalltau}
is consistent with the general case in \eqref{eq:SFF-small-tau}.
 
\section{Computation of SFF}\label{sec:SFF}

In this section we formulate a systematic method of computing
the small $\hbar$ expansion of the $\tau$-scaling limit of the SFF
in the general background $\{t_k\}$ of topological gravity.
In section~\ref{sec:SFFdiffeq}
we will first derive a key relation
\begin{align}
\partial_\tau\partial_0\text{SFF}
 &=\frac{\ri}{\hbar}\left[
  W_1\left(\beta+\frac{\ri\tau}{\hbar}\right)
  \partial_0 W_1\left(\beta-\frac{\ri\tau}{\hbar}\right)
 -\partial_0 W_1\left(\beta+\frac{\ri\tau}{\hbar}\right)
  W_1\left(\beta-\frac{\ri\tau}{\hbar}\right)
  \right],
\label{eq:dtauW2}
\end{align}
which relates SFF with (the $t_0$-derivative of) the one-point function
\begin{align}
W_1\left(\beta\pm\frac{\ri\tau}{\hbar}\right)
=\hbar\partial_0
 \left\langle Z\left(\beta\pm\frac{\ri\tau}{\hbar}\right)\right\rangle.
\end{align}
Here $\partial_k:=\partial_{t_k}$.
The relation \eqref{eq:dtauW2} enables us to compute
the small $\hbar$ expansion of the SFF from that of
the one-point function $W_1(\beta\pm\tfrac{\ri\tau}{\hbar})$.
As we will describe in section~\ref{sec:Gexp},
the small $\hbar$ expansion of
$W_1(\beta\pm\tfrac{\ri\tau}{\hbar})$
can be obtained by the method developed in \cite{Okuyama:2021cub}
with a slight modification. This is based on the KdV equation.
We will then explain in section~\ref{sec:SFFexp} how to
integrate $\partial_\tau\partial_0\text{SFF}$
to obtain the small $\hbar$ expansion of the SFF.

\subsection{Derivation of the key differential equation}
\label{sec:SFFdiffeq}

In this subsection we will derive \eqref{eq:dtauW2}.
As we have already seen in the previous sections,
the one- and two-point functions can be expressed in terms of
the CD kernel as \cite{Okuyama:2020ncd}
\begin{align}
\begin{aligned}
\langle Z(\bt)\rangle &=\int dE e^{-\bt E}K(E,E),\\
\langle Z(\bt_1)Z(\bt_2)\rangle_\conn
 &=\langle Z(\bt_1+\bt_2)\rangle
  -\int dE_1dE_2e^{-\bt_1E_1-\bt_2E_2}K(E_1,E_2)^2.
\end{aligned} 
\end{align}
The CD kernel is written as
\begin{equation}
 K(E_1,E_2)
 =\frac{\hbar\partial_0\psi(E_1)\psi(E_2)
       -\hbar\partial_0\psi(E_2)\psi(E_1)}{-E_1+E_2},
\label{eq:CDkernel}
\end{equation}
where $\psi(E)$ is the Baker-Akhiezer function. It satisfies
the Schr\"{o}dinger equation
\begin{equation}
-(\hbar^2\del_0^2+u)\psi(E)=E\psi(E).
\end{equation}
By using this equation it immediately follows from \eqref{eq:CDkernel}
that
\begin{equation}
 \hbar\del_0K(E_1,E_2)=\psi(E_1)\psi(E_2).
\end{equation}
Then we have
\begin{equation}
\begin{aligned}
\hbar\del_0\langle Z(\bt)\rangle
 &=W_1(\bt)=\int dE e^{-\bt E}\psi(E)^2,\\
\hbar\del_0\langle Z(\bt_1)Z(\bt_2)\rangle_\conn
 &=W_1(\bt_1+\bt_2)-\int dE_1dE_2e^{-\bt_1E_1-\bt_2E_2}
   2\psi(E_1)\psi(E_2)K(E_1,E_2).
\end{aligned} 
\end{equation}
When $\bt_{1,2}=\bt\pm\ri \tau\hbar^{-1}$, we find
\begin{equation}
\begin{aligned}
&\partial_\tau\del_0\langle Z(\bt_1)Z(\bt_2)\rangle_\conn\\
&=-\frac{\ri}{\hbar^2}
 \int dE_1dE_2e^{-\bt_1E_1-\bt_2E_2}
 2\psi(E_1)\psi(E_2)(-E_1+E_2)K(E_1,E_2)\\
&=-\frac{\ri}{\hbar^2}
 \int dE_1dE_2e^{-\bt_1E_1-\bt_2E_2}2\psi(E_1)\psi(E_2)
\left[\hbar\partial_0\psi(E_1)\psi(E_2)
     -\hbar\partial_0\psi(E_2)\psi(E_1)\right]\\
&=-\frac{\ri}{\hbar}\int dE_1dE_2e^{-\bt_1E_1-\bt_2E_2}
 \left[\partial_0\psi(E_1)^2\psi(E_2)^2
      -\psi(E_1)^2\partial_0\psi(E_2)^2\right]\\
&=-\frac{\ri}{\hbar}\left[\partial_0 W_1(\bt_1)W_1(\bt_2)
           -W_1(\bt_1)\partial_0 W_1(\bt_2)\right].
\end{aligned}
\label{eq:w2w1}
\end{equation}
Thus we have derived \eqref{eq:dtauW2}.

\subsection{Small $\hbar$ expansion of one-point function}
\label{sec:Gexp}

Let us next consider the small $\hbar$ expansion of
$W_1\left(\beta\pm\frac{\ri\tau}{\hbar}\right)$.
We can restrict ourselves to the case of
$W_1\left(\beta+\frac{\ri\tau}{\hbar}\right)$
without loss of generality,
as $W_1\left(\beta-\frac{\ri\tau}{\hbar}\right)$
is immediately obtained by complex conjugation.
The expansion can be done in two ways.
One way, which is easier to understand, is to use the results
of the 't Hooft expansion computed in \cite{Okuyama:2021cub}:
\begin{align}
W_1\left(\beta\right)
=\exp\left[\sum_{n=0}^\infty\hbar^{n-1}\tG_n(\lambda)\right],
\quad\lambda=\hbar\beta\ \mbox{fixed.}
\label{eq:W1tHooft}
\end{align}
Given this expansion, the small $\hbar$ expansion of
$W_1\left(\beta+\frac{\ri\tau}{\hbar}\right)$
is obtained by merely substituting $\lambda=\ri\tau+\beta\hbar$
into the expansion
and then re-expanding it in $\hbar$ as
\begin{align}
\begin{aligned}
W_1\left(\beta+\frac{\ri\tau}{\hbar}\right)
&=\exp\biggl[
  \hbar^{-1}\tG_0(\ri\tau)
 +\left(\tG_1(\ri\tau)+\tG_0'(\ri\tau)\beta\right)\\
&\hspace{4em}
 +\hbar\left(\tG_2(\ri\tau)+\tG_1'(\ri\tau)\beta
   +\frac{1}{2}\tG_0''(\ri\tau)\beta^2\right)
 +{\cal O}(\hbar^2)
  \biggr].
\end{aligned}
\end{align}
More specifically,
the first few $\tG_n$ are given by
\cite{Okuyama:2021cub}\footnote{
The elements here and there are identified as
$\tG_n=2^{(n-1)/2}G_n$,
$E_0=-u_0$,
$s^{\text{here}}=t^{\text{there}}=1-I_1$,
$E(\lambda)=-\xi_*$,
$E^{(j)}=-2^{-n/2}\xi_*^{(n)}$,
$z(\lambda)=-2^{-1/2}\ri z_*$ and
$\lambda=2^{1/2}s^{\text{there}}$.
The normalization of $W_1$ here
(i.e.~the constant part of $\tG_1$)
is also adjusted accordingly.
}
\begin{align}
\begin{aligned}
\tG_0
&=-E_0\lambda
 +\sum_{\substack{j_a\ge 0\\[.5ex]
                   \sum_a j_a=k\\[.5ex]
                   \sum_a aj_a=n}}
 \frac{(2n+k+1)!}{(2n+3)!}\frac{\lambda^{2n+3}}{2^{n+1}s^{2n+k+2}}
 \prod_{a=1}^\infty\frac{I_{a+1}^{j_a}}{j_a!(2a+1)!!^{j_a}}\\
&=-E_0\lambda
 +\frac{1}{12s^2}\lambda^3
 +\frac{I_2}{60s^5}\lambda^5
 +\left(\frac{I_2^2}{144s^8}+\frac{I_3}{840s^7}\right)\lambda^7
 +{\cal O}(\lambda^9),\\[1ex]
\tG_1&=
  \frac{1}{2}\log\left[\frac{\hbar}{8\pi}
  \frac{\partial_\lambda E(\lambda)}{E(\lambda)-E_0}\right],\\
\tG_2&=
  \frac{\ri I_2}{12s^2z(\lambda)}
 -\frac{5\ri}{24sz(\lambda)^3}
 -\frac{3E^{(1)}}{8z(\lambda)^4}
 +\frac{E^{(2)}}{4z(\lambda)^2E^{(1)}}
 -\frac{E^{(3)}}{8\left(E^{(1)}\right)^2}
 +\frac{\left(E^{(2)}\right)^2}{6\left(E^{(1)}\right)^3},
\end{aligned}
\label{eq:G0Etau}
\end{align}
where
\begin{align}
E(\lambda)
=-\partial_\lambda \tG_0,\qquad
z(\lambda)=\sqrt{E(\lambda)-E_0},\qquad
E^{(j)}=\partial_\lambda^jE(\lambda).
\label{eq:Eofla}
\end{align}
$I_k$ and $s=1-I_1$ were defined in section~\ref{sec:review}.
From this one obtains
\begin{align}
W_1\left(\beta+\frac{\ri\tau}{\hbar}\right)
=e^{G\left(\beta+\frac{\ri\tau}{\hbar}\right)}
=\exp\left[\sum_{n=0}^\infty\hbar^{n-1}G_n(\beta,\tau,\{I_k\})\right]
\label{eq:W1exp}
\end{align}
with
\begin{align}
\begin{aligned}
G_0&=\tG_0(\ri\tau),\\
G_1&=\frac{1}{2}\log\left[\frac{\hbar}{8\pi\ri}
     \frac{\partial_\tau E_\tau}{E_\tau-E_0}\right]
     -\beta E_\tau,\\
G_2&=\ri\left[
  \frac{I_2}{12s^2z_\tau}
 -\frac{5}{24sz_\tau^3}
 +\frac{3e(1)}{8z_\tau^4}
 -\frac{e(2)}{4z_\tau^2e(1)}
 +\frac{e(3)}{8e(1)^2}
 -\frac{e(2)^2}{6e(1)^3}\right.\\
&\hspace{2em}\left.
 +\left(\frac{e(1)}{2z_\tau^2}-\frac{e(2)}{2e(1)}\right)\beta
 +\frac{e(1)}{2}\beta^2
 \right].
\end{aligned}
\label{eq:G012}
\end{align}
Here $E_\tau=E(\ri\tau)$, $z_\tau=\sqrt{E_\tau-E_0}=z(\ri\tau)$
and $e(j)=\partial_\tau^j E_\tau=\ri^j E^{(j)}(\ri\tau)$.

The other way to compute the small $\hbar$ expansion of
$W_1\left(\beta+\frac{\ri\tau}{\hbar}\right)$,
which is technically more efficient,
is to solve the KdV equation directly
with the expansion \eqref{eq:W1exp}.
This can be done by the method of \cite{Okuyama:2021cub}
with only a slight modification.
In what follows let us describe the method
in our present notation.
First, recall that
$W_1=W_1\left(\beta+\frac{\ri\tau}{\hbar}\right)$
satisfies the KdV equation \cite{Okuyama:2019xbv}
\begin{align}
\partial_1 W_1=u\partial_0W_1+\frac{\hbar^2}{6}\partial_0^3W_1.
\label{eq:W1KdV}
\end{align}
In terms of $G=\log W_1$, this is written as
\begin{align}
\partial_1 G
 =u\partial_0 G
 +\frac{\hbar^2}{6}\Bigl(\partial_0^3 G+3\partial_0 G\partial_0^2 G
   +(\partial_0 G)^3 \Bigr).
\label{eq:Gdiffeq}
\end{align}
Here, $u$ is the specific heat
of the general topological gravity.
It obeys the KdV equation and its genus expansion
\begin{align}
u=\sum_{g=0}^\infty 2^gu_g\hbar^{2g}
\label{eq:uexp}
\end{align}
was computed by means of a recurrence relation.\footnote{See,
e.g.~\cite{Okuyama:2019xbv}
with the identification
$t^{\text{there}}=s^{\text{here}},\ B_n=(-1)^{n+1}I_{n+1}\ (n\ge 1)$.}
Note, in particular, that the leading order part is equal to
the threshold energy 
\begin{align}
u_0=-E_0.
\end{align}
By plugging the expansions \eqref{eq:uexp} and
\begin{align}
G=\sum_{n=0}^\infty\hbar^{n-1}G_n
\label{eq:Gexp}
\end{align}
into \eqref{eq:Gdiffeq},
one obtains the recurrence relation
\begin{align}
\label{eq:Grec0}
-\partial_s G_0&=\frac{1}{6}(\partial_0 G_0)^3,\\
\label{eq:Grec1}
D G_1&=\frac{1}{2}\partial_0 G_0\partial_0^2 G_0
\end{align}
for $n=0,1$ and
\begin{align}
\begin{aligned}
D G_n
  =\frac{1}{6}\sum_{\substack{0\le i,j,k<n\\[1ex] i+j+k=n}}
     \partial_0 G_i \partial_0 G_j \partial_0 G_k
  +\frac{1}{2}\sum_{k=0}^{n-1}\partial_0 G_{n-k-1} \partial_0^2 G_{k}
  +\frac{1}{6}\partial_0^3 G_{n-2}
  +\sum_{k=1}^{\lfloor\frac{n}{2}\rfloor}2^ku_k\partial_0 G_{n-2k}
\end{aligned}
\label{eq:Grecgen}
\end{align}
for $n\ge 2$.\footnote{$G_n$ here is related with $G_n$
in \cite{Okuyama:2021cub} as
$G_n^{\text{here}}(\beta,\tau=0,\{I_k\})
 =2^{(n-1)/2}G_n^{\text{there}}(s=2^{-1/2}\hbar\beta,\{I_k\})$,
up to the constant part of $G_1$.}
Here we have introduced the differential operator
\begin{align}
D := -\partial_s-\frac{1}{2}(\partial_0 G_0)^2\partial_0.
\label{eq:Ddef}
\end{align}
Note also that instead of $t_0,t_1$ we regard $s$ and $E_0$
as independent variables and treat $t_{k\ge 2}$ as parameters
\cite{Okuyama:2019xbv}. From this viewpoint $\partial_{0,1}$
are understood as
\begin{align}
\partial_0
 =-\frac{1}{s}\left(\partial_{E_0}+I_2\partial_s\right),\qquad
\partial_1
 =-E_0\partial_0-\partial_s.
\label{eq:d0d1}
\end{align}

To solve the recurrence relation
\eqref{eq:Grec0}--\eqref{eq:Grecgen},
it is helpful to recall some relevant formulas
(see \cite{Okuyama:2021cub}):
\begin{align}
\begin{aligned}
\partial_0 s
 &=-\frac{I_2}{s},\quad&
\partial_0 I_n
 &=\frac{I_{n+1}}{s}\quad (n\ge 1),\ \\
\partial_0G_0
 &=2\ri z_\tau,&
\partial_0 z_\tau
 &=\frac{1}{2sz_\tau}-\frac{\partial_\tau z_\tau}{z_\tau},&
\partial_0 e(j)
 &=-2\partial_\tau^{j+1}z_\tau,\\
\partial_s z_\tau
 &=-e(1).
\end{aligned}
\end{align}
The differential operator \eqref{eq:Ddef} is then written as
\begin{align}
D=-\partial_s+2 z_\tau^2\partial_0.
\end{align}
It has the properties
\begin{align}
\label{eq:Dprop1}
Dz_\tau&=\frac{z_\tau}{s},\qquad
D E_\tau=0,\\
\label{eq:Dprop2}
De(n)
&=8\sum_{k=0}^{n-1}\sum_{\ell=0}^k\sum_{m=0}^\ell
   \frac{k!}{(k-\ell)!(\ell-m)!m!}
   \partial_\tau^{(n-\ell)}z_\tau\,
   \partial_\tau^{(\ell-m)}z_\tau\,
   \partial_\tau^{(m+1)}z_\tau.
\end{align}
As explained in \cite{Okuyama:2020vrh},
$\partial_\tau^n z_\tau$ can be expressed
in terms of $e(j)$ and $z_\tau$:
\begin{align}
\begin{aligned}
\partial_\tau z_\tau
 &=\frac{e(1)}{2z_\tau},\\
\partial_\tau^2 z_\tau
 &=\frac{e(2)}{2z_\tau}-\frac{e(1)^2}{4z_\tau^3},\\
\partial_\tau^3 z_\tau
 &=\frac{e(3)}{2z_\tau}-\frac{3e(1)e(2)}{4z_\tau^3}
   +\frac{3e(1)^3}{8z_\tau^5}.
\end{aligned}
\end{align}

We are now in a position to solve the recurrence relation.
In \cite{Okuyama:2021cub} the recurrence relation
\eqref{eq:Grec0}--\eqref{eq:Grecgen}
was solved with the initial data
\begin{align}
\begin{aligned}
G_0^{\text{before}}
  =\tG_0(\ri\tau),\qquad
G_1^{\text{before}}
  =\frac{1}{2}\log\left[\frac{\hbar}{8\pi\ri}
   \frac{\partial_\tau E_\tau}{E_\tau-E_0}\right].
\end{aligned}
\label{eq:oldG01}
\end{align}
Here, we would like to solve the same recurrence relation
with the initial data given in \eqref{eq:G012}.
The only difference from \cite{Okuyama:2021cub}
is that here $G_1$ has an extra $\beta$-dependent term
$-\beta E_\tau$.
It is clear from \eqref{eq:Dprop1} that
\eqref{eq:Grec1} is satisfied by
$G_1$ with this term,
given that \eqref{eq:Grec1} is satisfied by $G_1^{\text{before}}$.

To solve the recurrence relation, we can use almost the same algorithm
we developed in \cite{Okuyama:2021cub}.
The only difference is the step (v),
which is modified accordingly
so that it works for the $\beta$-dependent parts as well.
The modified algorithm is as follows:

\renewcommand{\theenumi}{\roman{enumi}}
\renewcommand{\labelenumi}{(\theenumi)}
\begin{enumerate}
\item Compute the r.h.s.~of \eqref{eq:Grecgen} and express it as
a polynomial in the variables  $s^{-1}$, $I_{k\ge 2}$,
$z_\tau^{-1}$, $e(1)^{-1}$ and $e(n),\ {n\ge 1}$.

\item Let $s^{-m}f\bigl(I_k,z_\tau,e(n)\bigr)$ denote the 
highest-order part in $s^{-1}$ of the obtained expression. This part
can arise only from
\begin{align}
D\left(
  \frac{f\bigl(I_k,z_\tau,e(n)\bigr)}
       {2(m-2)s^{m-2}z_\tau^2 I_2}\right).
\end{align}
Therefore subtract this from the obtained expression.

\item Repeat procedure (ii) down to $m=3$. Then all the terms of
order $s^{-2}$ automatically disappear and the remaining terms
are of order $s^{-1}$ or $s^{0}$. Note also that the expression
does not contain any $I_k$.

\item In the result of (iii), collect all the terms of order $s^{-1}$ 
and let $s^{-1}z_\tau\partial_{z_\tau}g\bigl(z_\tau,e(n)\bigr)$
denote the sum of them. This part arises from
\begin{align}
D g\bigl(z_\tau,e(n)\bigr).
\end{align}
Therefore subtract this from the result of (iii). The remainder
turns out to be independent of $s$.

\item
In the obtained expression, let
\begin{align}
\frac{e(1)^m h\bigl(e(n\ge 2)\bigr)}{z_\tau}
\end{align}
denote the part which is of the order $z_\tau^{-1}$ as well as of
the highest order in $e(1)$. This part arises from
\begin{align}
D\left[
-\frac{1}{2e(1)}\int^{\ta} da
\left(a^{-1}e(1)\right)^m h\bigl(e(n)\to a^{-n-1}e(n)\bigr)
\right]_{\ta=1}.
\end{align}
(Here, $a$ and $\ta$ are intermediate formal variables.)
Therefore subtract this from the obtained expression.

\item Repeat procedure (v) until the resulting expression vanishes.

\item By summing up all the above-obtained primitive functions, we
obtain $G_n$.

\end{enumerate}
Using the above algorithm we have computed $G_n$ for $n\le 12$.
For instance, $G_3$ is computed as
\begin{align}
\begin{aligned}
G_3&=
-\frac{5}{32s^2\zt^6}
+\frac{I_2}{8s^3\zt^4}
-\frac{I_3}{24s^3\zt^2}
-\frac{I_2^2}{12s^4\zt^2}
-\frac{5I_2e(1)}{96s^2\zt^5}
-\frac{3e(1)^2}{4\zt^8}
+\frac{35e(1)}{64s\zt^7}
+\frac{11e(2)}{16\zt^6}\\
&\hspace{1em}
-\frac{3e(3)}{16\zt^4e(1)}
-\frac{5e(2)}{32s\zt^5e(1)}
+\frac{I_2e(2)}{48s^2\zt^3e(1)}
+\frac{e(4)}{16\zt^2e(1)^2}
+\frac{e(2)^2}{8\zt^4e(1)^2}
-\frac{e(5)}{48e(1)^3}\\
&\hspace{1em}
-\frac{7e(2)e(3)}{24\zt^2e(1)^3}
+\frac{e(3)^2}{8e(1)^4}
+\frac{e(2)^3}{4\zt^2e(1)^4}
+\frac{e(2)e(4)}{6e(1)^4}
-\frac{3e(2)^2e(3)}{4e(1)^5}
+\frac{e(2)^4}{2e(1)^6}
\\
&\hspace{1em}
+\left(
 \frac{5e(1)}{16s\zt^5}
-\frac{3e(1)^2}{4\zt^6}
+\frac{5e(2)}{8\zt^4}
-\frac{I_2e(1)}{24s^2\zt^3}
-\frac{e(3)}{4\zt^2e(1)}
+\frac{e(4)}{8e(1)^2}
+\frac{e(2)^2}{4\zt^2e(1)^2}\right.\\
&\hspace{2.8em}\left.
-\frac{7e(2)e(3)}{12e(1)^3}
+\frac{e(2)^3}{2e(1)^4}
\right)\beta
+\left(-\frac{e(1)^2}{4\zt^4}
+\frac{e(2)}{4\zt^2}
+\frac{e(2)^2}{4e(1)^2}
-\frac{e(3)}{4e(1)}\right)\beta^2
+\frac{e(2)}{6}\beta^3.
\end{aligned}
\end{align}

Let us finally comment on the small $\hbar$ expansion of
$W_1\left(\beta-\frac{\ri\tau}{\hbar}\right)$.
As we mentioned already, it is obtained
from \eqref{eq:W1exp} by complex conjugation:
\begin{align}
W_1\left(\beta-\frac{\ri\tau}{\hbar}\right)
=\exp\left[\sum_{n=0}^\infty\hbar^{n-1}
 \overline{G_n}\right].
\label{eq:bW1exp}
\end{align}
Furthermore, $\overline{G_n}$ is related to $G_n$ by
\begin{align}
\overline{G_n}=(-1)^{n+1}G_n+\frac{\pi\ri}{2}\delta_{n,1}.
\label{eq:GbGrel}
\end{align}
This can be shown directly for $n=0,1$ and
inductively for $n\ge 2$
by using the recurrence relation \eqref{eq:Grecgen}.

\subsection{Small $\hbar$ expansion of SFF}
\label{sec:SFFexp}

Using the results obtained above,
one can compute the small $\hbar$ expansion of the SFF.
Substituting \eqref{eq:W1exp}, \eqref{eq:bW1exp}
and the explicit form of $G_1$ in \eqref{eq:G012}
into \eqref{eq:dtauW2}, one obtains
\begin{align}
\begin{aligned}
\partial_\tau\partial_0\text{SFF}
 &=\frac{\ri}{\hbar}(\partial_0 \overline{G} -\partial_0 G)
   e^{G+\overline{G}}\\
 &=-\frac{2\ri}{\hbar}\sum_{k=0}^\infty\hbar^{2k-1}\partial_0 G_{2k}
   \exp\left(G_1+\overline{G_1}
            +2\sum_{k=1}^\infty \hbar^{2k}G_{2k+1}\right)\\
 &=\frac{e^{-2\beta E_\tau}}{4\pi\ri}
   \frac{\partial_\tau E_\tau}{E_\tau-E_0}
   \sum_{k=0}^\infty\hbar^{2k-1}\partial_0 G_{2k}
   \exp\left(\sum_{k=1}^\infty 2G_{2k+1}\hbar^{2k}\right)\\
 &=:\frac{e^{-2\beta E_\tau}}{2\pi\hbar}\sum_{n=0}^\infty
    c_n\hbar^{2n},
\end{aligned}
\label{eq:d0dSFFexp}
\end{align}
where the first two of $c_n$ are computed as
\begin{align}
\begin{aligned}
c_0&=\frac{e(1)}{z_\tau},\\
c_1&=
\frac{e(1)e(2)}{3z_\tau}\beta^3
+\left(
 -\frac{3e(1)^3}{8z_\tau^5}
 +\frac{e(1)e(2)}{4z_\tau^3}
 -\frac{e(3)}{2z_\tau}
 +\frac{e(2)^2}{2z_\tau e(1)}
 \right)\beta^2\\
&\hspace{1em}
+\left(
 -\frac{15e(1)^3}{16z_\tau^7}
 +\frac{3e(1)e(2)}{4z_\tau^5}
 -\frac{e(3)}{4z_\tau^3}
 +\frac{3e(1)^2}{8sz_\tau^6}
 -\frac{I_2e(1)^2}{12s^2z_\tau^4}\right.\\
&\hspace{2.8em}\left.
 +\frac{e(4)}{4z_\tau e(1)}
 +\frac{e(2)^2}{4z_\tau^3e(1)}
 -\frac{7e(2)e(3)}{6z_\tau e(1)^2}
 +\frac{e(2)^3}{z_\tau e(1)^3}
\right)\beta\\
&\hspace{1em}
 -\frac{5e(1)}{32s^2z_\tau^7}
 +\frac{25e(1)e(2)}{32z_\tau^7}
 -\frac{3e(3)}{16z_\tau^5}
 +\frac{9e(1)^2}{16sz_\tau^8}
 +\frac{I_2e(1)}{8s^3z_\tau^5}
 -\frac{I_2^2e(1)}{12s^4z_\tau^3}
 -\frac{I_2e(1)^2}{12s^2z_\tau^6}
 -\frac{3e(2)}{16sz_\tau^6}\\
&\hspace{1em}
 -\frac{I_3e(1)}{24s^3z_\tau^3}
 +\frac{I_2e(2)}{24s^2z_\tau^4}
 -\frac{105e(1)^3}{128z_\tau^9}
 +\frac{3e(2)^2}{32z_\tau^5e(1)}
 +\frac{e(4)}{16z_\tau^3e(1)}
 -\frac{e(5)}{24z_\tau e(1)^2}
 -\frac{7e(2)e(3)}{24z_\tau^3e(1)^2}\\
&\hspace{1em}
 +\frac{e(2)^3}{4z_\tau^3e(1)^3}
 +\frac{e(3)^2}{4z_\tau e(1)^3}
 +\frac{e(2)e(4)}{3z_\tau e(1)^3}
 -\frac{3e(2)^2e(3)}{2z_\tau e(1)^4}
 +\frac{e(2)^4}{z_\tau e(1)^5}.
\end{aligned}
\label{eq:c01}
\end{align}

Let us next consider the expansion
\begin{align}
\partial_\tau\text{SFF}
 =\frac{e^{-2\beta E_\tau}}{2\pi\hbar}
  \sum_{n=0}^\infty b_n\hbar^{2n}.
\label{eq:dSFFexp}
\end{align}
Comparing \eqref{eq:dSFFexp} with \eqref{eq:d0dSFFexp},
we see that $b_n$ and $c_n$ are related as
\begin{align}
\begin{aligned}
c_n
 &=e^{2\beta E_\tau}\partial_0 e^{-2\beta E_\tau}b_n\\
 &=\left(\partial_0-2\beta\partial_0 E_\tau\right)b_n\\
 &=\left(\partial_0+2\beta\frac{e(1)}{z_\tau}\right)b_n.
\end{aligned}
\label{eq:bncnrel}
\end{align}
For $n=0$, we immediately see that $b_0=1/(2\beta)$
gives $c_0$ in \eqref{eq:c01}.
For $n\ge 1$, let us assume that $b_n$ is a polynomial in $\beta$.
Then, the highest-order term of this polynomial
is determined as that of $c_n$ divided by $2\beta e(1)/z_\tau$.
Subtracting this contribution from both sides of \eqref{eq:bncnrel},
one can determine the next to the highest-order term
in the same manner.
In this way, one can determine $b_n$ as a polynomial of $\beta$
without integrating in $t_0$.
For instance, the first two of them are
\begin{align}
\begin{aligned}
b_0&=\frac{1}{2\beta},\\
b_1&=
 \frac{e(2)}{6}\beta^2
 +\left(
 -\frac{e(1)^2}{8z_\tau^4}
 -\frac{e(3)}{6e(1)}
 +\frac{e(2)^2}{4e(1)^2}
  \right)\beta\\
&\hspace{1em}
 -\frac{I_2e(1)}{24s^2z_\tau^3}
 +\frac{e(2)}{16z_\tau^4}
 +\frac{e(1)}{16sz_\tau^5}
 -\frac{e(1)^2}{8z_\tau^6}
 +\frac{e(4)}{24e(1)^2}
 -\frac{e(2)e(3)}{4e(1)^3}
 +\frac{e(2)^3}{4e(1)^4}.
\end{aligned}
\label{eq:bn}
\end{align}

Let us finally consider the small $\hbar$ expansion \eqref{eq:SFFg-def}
of the SFF:
\begin{align}
\begin{aligned}
\text{SFF}
&=\text{SFF}_{g=0}+\text{SFF}_{g\ge 1}\\
&=\hbar^{-1}\text{SFF}_0+\sum_{g=1}^\infty\hbar^{2g-1}\text{SFF}_g.
\end{aligned}
\end{align}
The leading term $\text{SFF}_0$ is given in \eqref{eq:SFF0}.
It is easy to verify that the $\tau$-derivative of \eqref{eq:SFF0}
reproduces the leading order part of \eqref{eq:dSFFexp}
with $b_0=1/(2\beta)$.
Here, we are interested in the higher order
part $\text{SFF}_{g\ge 1}$.
As explained in section~\ref{sec:summary}
(see, in particular \eqref{eq:ZErf-cg}),
a key feature of $\text{SFF}_{g\ge 1}$
is that it takes the form
\begin{align}
\begin{aligned}
\mathrm{SFF}_{g\ge 1}
 &=\Erf\left(\sqrt{2\beta}z_\tau\right)
 \langle Z(2\beta)\rangle_{g\ge 1}
  +\frac{e^{-2\beta E_\tau}}{2\pi\hbar}
   \sum_{g=1}^\infty a_g\hbar^{2g}
\end{aligned}
\label{eq:SFFexp1}
\end{align}
with $a_g$ being polynomial in $\beta$.
The small $\hbar$ expansion of the one-point function
$\langle Z(2\beta)\rangle$
was computed in \cite{Okuyama:2019xbv}:
\begin{align}
\langle Z(2\beta)\rangle_{g\ge 1}
 =\frac{e^{-2\beta E_0}}{\sqrt{2\pi(2\beta)^3}\sqrt{2}\hbar}
\sum_{g=1}^\infty\hbar^{2g} 2^g Z_g(2\beta),
\label{eq:Zexp}
\end{align}
where
\begin{align}
\begin{aligned}
Z_1(\beta)&=
 \frac{1}{24s}\beta^3
 +\frac{I_2}{24s^2}\beta^2,\\
Z_2(\beta)&=
 \frac{1}{1152s^3}\beta^6
 +\frac{29I_2}{5760s^4}\beta^5
+\left(
 \frac{7I_2^2}{480s^5}
 +\frac{29I_3}{5760s^4}
 \right)\beta^4
+\left(
 \frac{7I_2^3}{288s^6}
 +\frac{29I_2I_3}{1440s^5}
 +\frac{I_4}{384s^4}
 \right)\beta^3\\
&\hspace{1em}
+\left(
 \frac{7I_2^4}{288s^7}
 +\frac{5I_2^2I_3}{144s^6}
 +\frac{29I_3^2}{5760s^5}
 +\frac{11I_2I_4}{1440s^5}
 +\frac{I_5}{1152s^4}
 \right)\beta^2.
\end{aligned}
\end{align}
Comparing \eqref{eq:SFFexp1} with \eqref{eq:dSFFexp},
we see that $a_g$ and $b_g$ are related as
\begin{align}
b_g
 =\frac{e(1)}{2\beta z_\tau}2^g Z_g(2\beta)
 +\bigl(\partial_\tau-2\beta e(1)\bigr)a_g.
\end{align}
Given that $a_g$ are polynomials in $\beta$,
we can calculate them from $b_g$ without integrating in $\tau$,
in essentially the same way as we obtained $b_g$ from $c_g$.
For instance, $a_1$ is obtained as
\begin{align}
a_1&=
\left(
  \frac{1}{6sz_\tau}
 -\frac{e(2)}{12e(1)}
\right)\beta
 +\frac{e(1)}{16z_\tau^4}
 +\frac{I_2}{12s^2z_\tau}
 -\frac{1}{24sz_\tau^3}
 +\frac{e(3)}{24e(1)^2}
 -\frac{e(2)^2}{12e(1)^3}.
\end{align}
One sees that 
the terms which do not contain $e(j)$
precisely give $d_1$ in \eqref{eq:nonErf1};
the other terms comprise
$f_1$ given by \eqref{eq:fg-str} with \eqref{eq:h1n}.
The expression of $a_2$ is already rather long
and we relegate it to Appendix~\ref{app:a2}.
We have computed $a_g$ for $g\le 5$.

\section{Higher order corrections to CD kernel}\label{sec:CD}

In this section we consider the higher order correction of
the CD kernel beyond the sine kernel approximation.
That is, we consider a small $\hbar$ expansion
of the CD kernel.
In principle, this is not a difficult task since
the CD kernel is expressed as a bilinear form \eqref{eq:CDkernel}
in terms of the BA function $\psi(E)$ whose $\hbar$-expansion
was calculated \cite{Okuyama:2020ncd}.
The purpose of this section is to present
a concrete, technically efficient method.

\subsection{Small $\hbar$ expansion of BA function}

To compute the expansion of the CD kernel,
one first needs to compute the small $\hbar$ expansion of
$\psi(E+\frac{1}{2}\hbar\omega)$ rather than $\psi(E)$.
Of course, this is obtained by re-expanding
the results of \cite{Okuyama:2020ncd}
with the replacement $E\to E+\frac{1}{2}\hbar\omega$.
For the sake of technical efficiency,
however, here we compute the expansion
\begin{align}
\log\psi(E+\tfrac{1}{2}\hbar\omega)
 =A(E+\tfrac{1}{2}\hbar\omega)
 =\sum_{n=0}^\infty\hbar^{n-1}A_n(E,\omega)
\end{align}
directly by means of a recurrence relation.

It was derived in \cite{Okuyama:2020ncd} that
\begin{align}
v(E):=\hbar\partial_0 A(E)
\end{align}
satisfies the equation
\begin{align}
\begin{aligned}
v^2+\hbar\partial_0 v
 &=-E-u.
\end{aligned}
\label{eq:vdiffeq}
\end{align}
Here, $u$ is the specific heat of the general topological gravity
and we again assume that its genus expansion \eqref{eq:uexp}
is given.
Since we want to compute the $\hbar$ expansion of 
$\psi$ with a shifted argument,
we replace $E$ in \eqref{eq:vdiffeq} by $E+\tfrac{1}{2}\hbar\omega$
and plug the expansion of the form
\begin{align}
v=v(E+\tfrac{1}{2}\hbar\omega)
 =\sum_{n=0}^\infty\hbar^{n}v_n(E,\omega)
\end{align}
into \eqref{eq:vdiffeq}.
Expanding both sides of the equation in $\hbar$,
we obtain the recurrence relation
\begin{align}
\begin{aligned}
v_0^2&=-E+E_0,\qquad
2v_0v_1+\partial_0v_0=-\frac{\omega}{2},\\
v_n
&=-\frac{1}{2v_0}
   \left(\partial_0 v_{n-1}+\sum_{k=1}^{n-1}v_k v_{n-k}
        +\left\{\begin{array}{ll}
2^{\frac{n}{2}}u_{\frac{n}{2}}&\mbox{($n$ even)}\\
                           0&\mbox{($n$ odd)}
                 \end{array}\right.
   \right),\quad n\ge 2.
\end{aligned}
\label{eq:vrecrel}
\end{align}
The first two equations are solved as
\begin{align}
\begin{aligned}
v_0&=\ri z=\ri\sqrt{E-E_0},\\
v_1&=-\frac{1}{4sz^2}+\frac{\ri\omega}{4z}.
\end{aligned}
\label{eq:v01}
\end{align}
$v_n\ (n\ge 2)$ are also immediately obtained
by the recurrence relation given the form of $u_g$.
For instance, using
\begin{align}
u_1=\frac{I_2^2}{12s^4}+\frac{I_3}{24s^3}
\end{align}
one obtains
\begin{align}
\begin{aligned}
v_2&=
 \ri\left(
 \frac{I_2^2}{12zs^4}
 +\frac{I_3}{24zs^3}
 -\frac{I_2}{8z^3s^3}
 +\frac{5}{32z^5s^2}
 \right)
 +\frac{1}{8z^4s}\omega
 -\ri\frac{1}{32z^3}\omega^2,\\
v_3&=
 -\frac{I_2^3}{6s^6z^2}
 -\frac{7I_2I_3}{48s^5z^2}
 +\frac{11I_2^2}{48s^5z^4}
 -\frac{I_4}{48s^4z^2}
 +\frac{I_3}{12s^4z^4}
 -\frac{9I_2}{32s^4z^6}
 +\frac{15}{64s^3z^8}
\\
&\hspace{1em}
 +\ri\left(
 -\frac{I_2^2}{48s^4z^3}
 -\frac{I_3}{96s^3z^3}
 +\frac{3I_2}{32s^3z^5}
 -\frac{25}{128s^2z^7}
 \right)\omega
 -\frac{1}{16sz^6}\omega^2
 +\ri\frac{1}{128z^5}\omega^3.
\end{aligned}
\label{eq:vn}
\end{align}
Note that the form of the recurrence relation
\eqref{eq:vrecrel} for $v_{n\ge 2}$
is not affected by the shift $E\to E+\frac{1}{2}\hbar\omega$.
The $\omega$-dependence enters entirely through
the initial data, i.e.~the form of $v_1$.

One can easily obtain $A_n$ from $v_n$
through the relation $\partial_0 A_n=v_n$.
Due to the structure \eqref{eq:d0d1} of $\partial_0$,
the form of the $s$-dependent part of $A_n$ can be determined
without integrating in $t_0$.
(We use a similar logic as we obtained $b_n$ from $c_n$
in the last section.)
The $s$-independent part can also be determined
with the help of the formulas
(the proofs of which are straightforward)
\begin{align}
\begin{aligned}
\int^{t_0}dt_0\frac{1}{sz^2}&=2\log z,\\
\int^{t_0}dt_0\frac{1}{sz^{2k}}&=-\frac{1}{(k-1)z^{2(k-1)}}
\quad(k\ge 2),\\
\int^{t_0}dt_0\frac{1}{z^{2k-1}}
 &=\frac{\Gamma(\frac{3}{2}-k)}{\Gamma(\frac{1}{2})}
   \rho_0^{(k-1)}
  =\frac{(-2)^{k-1}}{(2k-3)!!}\rho_0^{(k-1)}\quad(k\ge 0).
\end{aligned}
\end{align}
Here
\begin{align}
\rho_0^{(n)}=\partial_E^n\rho_0(E)\quad (n\ge 0),\qquad
\rho_0^{(-1)}=\int_{E_0}^E dE\rho_0(E).
\end{align}
We thus obtain
\begin{align}
\begin{aligned}
A_0&=\frac{\ri}{2}\rho_0^{(-1)}
 =\frac{\ri}{2}\int_{E_0}^E dE\rho_0(E),\\
A_1&=-\frac{1}{2}\log z+\ri\frac{\rho_0}{4}\omega
  -\frac{1}{2}\log(4\pi),\\
A_2&=
 \ri\left(
  \frac{I_2}{24s^2z}
 -\frac{5}{48sz^3}\right)
 -\frac{1}{8z^2}\omega
 +\ri\frac{\rho_0^{(1)}}{16}\omega^2,\\
A_3&=
 -\frac{I_2^2}{24s^4z^2}
 -\frac{I_3}{48s^3z^2}
 +\frac{I_2}{16s^3z^4}
 -\frac{5}{64s^2z^6}
 +\ri\left(
  -\frac{I_2}{96s^2z^3}
  +\frac{5}{64sz^5}\right)\omega\\
&\hspace{1em}
 +\frac{1}{32z^4}\omega^2
 +\ri\frac{\rho_0^{(2)}}{96}\omega^3.
\end{aligned}
\label{eq:An}
\end{align}
The constant part of $A_1$ is determined
in accordance with the $\omega=0$ case \cite{Okuyama:2020ncd}.

\subsection{Small $\hbar$ expansion of CD kernel}

Let us now compute the small $\hbar$ expansion of the CD kernel.
We should first note that this is a WKB-type asymptotic expansion.
As is known \cite{Saad:2019lba},
there appear some qualitative differences
between the cases of $E>E_0$ and $E<E_0$.
In this paper we consider the case of $E>E_0$.
In this case the two independent
BA functions are given by
\begin{align}
\psi(E)\quad\mbox{and}\quad
\overline{\psi(E)}=\psi(E)\Big|_{z\to -z},
\end{align}
i.e.~they are complex conjugate to each other.
Due to this complex structure,
for $E>E_0$ there are two saddles which equally
contribute to the CD kernel \cite{Saad:2019lba}.
Therefore,
as long as the perturbative expansion in $\hbar$ concerns,
the CD kernel is approximated by
\begin{align}
K(E_1,E_2)=\tK(E_1,E_2)+\tK(E_2,E_1)
\end{align}
with
\begin{align}
\begin{aligned}
\tK(E_1,E_2)
 &=\frac{\hbar\partial_0\psi(E_1)\overline{\psi(E_2)}
        -\psi(E_1)\hbar\partial_0\overline{\psi(E_2)}}{-E_1+E_2}\\[1ex]
 &=-\frac{v(E_1)-\overline{v(E_2)}}{\hbar\omega}
   e^{A(E_1)+\overline{A(E_2)}}.
\end{aligned}
\label{eq:tK}
\end{align}

By using the recurrence relation \eqref{eq:vrecrel}
it is easy to prove that
\begin{align}
\overline{v_n(E,-\omega)}=(-1)^{n+1}v_n(E,\omega),\qquad
\overline{A_n(E,-\omega)}=(-1)^{n+1}A_n(E,\omega).
\end{align}
Therefore \eqref{eq:tK} is written as
\begin{align}
\tK(E_1,E_2)
=-\frac{1}{\hbar\omega}
 \left(\sum_{k=0}^\infty 2v_{2k}(E,\omega)\hbar^{2k}\right)
 \exp\left(\sum_{k=0}^\infty 2A_{2k+1}(E,\omega)\hbar^{2k}\right).
\end{align}
Similarly, we have
\begin{align}
\tK(E_2,E_1)
=-\frac{1}{\hbar\omega}
 \sum_{k=0}^\infty 2\overline{v_{2k}(E,\omega)}\hbar^{2k}
 \exp\left[\sum_{k=0}^\infty 2\overline{A_{2k+1}(E,\omega)}\hbar^{2k}
     \right].
\end{align}
Thus, the CD kernel is given by
\begin{align}
K(E_1,E_2)
 =\mathrm{Re}\left\{2\tK(E_1,E_2)\right\}.
\end{align}
Substituting the explicit form of $A_1(E,\omega)$ in \eqref{eq:An}
we obtain
\begin{align}
\begin{aligned}
2\pi K(E_1,E_2)
 &=\mathrm{Re}\left\{
 -\frac{2}{\hbar\omega z}
  \left(\sum_{k=0}^\infty v_{2k}(E,\omega)\hbar^{2k}\right)
  \exp\left(2\sum_{k=1}^\infty A_{2k+1}(E,\omega)\hbar^{2k}\right)
  e^{\ri\omega\rho_0/2}
 \right\}\\
 &=\mathrm{Re}\left\{X e^{\ri\omega\rho_0/2}\right\}\\
 &=\mathrm{Re}\{X\}\cos\frac{\omega\rho_0}{2}
  -\mathrm{Im}\{X\}\sin\frac{\omega\rho_0}{2},
\end{aligned}
\label{eq:CDexpform}
\end{align}
where
\begin{align}
X:= -\frac{2}{\hbar\omega z}
  \left(\sum_{k=0}^\infty v_{2k}(E,\omega)\hbar^{2k}\right)
  \exp\left(2\sum_{k=1}^\infty A_{2k+1}(E,\omega)\hbar^{2k}\right).
\label{eq:X}
\end{align}
With the forms of $v_n$ and $A_n$
obtained in \eqref{eq:v01}, \eqref{eq:vn} and \eqref{eq:An},
this gives the small $\hbar$ expansion of the CD kernel.

A few comments are in order.
First, substituting $v_0=\ri z$ one sees that
the expression \eqref{eq:CDexpform}
correctly reproduces the sine kernel \eqref{eq:sine-kernel}
at the order of $\hbar^{-1}$.
Second, recall that
the diagonal part of the CD kernel,
i.e.~\eqref{eq:CDexpform} in the limit of $\omega\to 0$,
is equal to the eigenvalue density \eqref{eq:K-diag}.
This is evident for the genus zero part
at the order of $\hbar^{-1}$,
but also implies that
\begin{align}
\sum_{g=1}^\infty\hbar^{2g-1}\rho_g(E)
 =\mathrm{Re}\{X\}\Big|_{\omega=0}.
\label{eq:rhoinX}
\end{align}
We checked that this indeed reproduces
$\rho_g$ in \eqref{eq:rho1and2}
which were
obtained from the result of $\langle Z(\beta)\rangle_g$
in \cite{Okuyama:2019xbv}.

By taking \eqref{eq:rhoinX} into account,
the expansion \eqref{eq:CDexpform} is written explicitly as
\begin{align}
\begin{aligned}
&2\pi K(E_1,E_2)\\
&=\frac{2}{\hbar\omega}
 \Biggl\{1+\frac{\omega^2}{32z^4}\hbar^2
 +\Biggl[
 +\left(
 -\frac{49I_2^2}{4608s^4z^6}
 -\frac{I_3}{192s^3z^6}
 +\frac{49I_2}{1536s^3z^8}
 -\frac{105}{2048s^2z^{10}}\right)\omega^2
\\
 &\hspace{4em}
+\left(
 \frac{I_2\rho_0^{(2)}}{2304s^2z^3}
 -\frac{\rho_0^{(2)}}{1536sz^5}
 +\frac{11}{2048z^8}
  \right)\omega^4
-\frac{(\rho_0^{(2)})^2}{4608}\omega^6
 \Biggr]\hbar^4
 +{\cal O}(\hbar^6)\Biggr\}
  \sin\frac{\omega\rho_0}{2}\\
&\hspace{1em}
 +\Biggl\{
  \left[\rho_1+\frac{\rho_0^{(2)}}{24}\omega^2\right]\hbar\\
&\hspace{1.8em}
 +\Biggl[\rho_2
 +\left(
       -\frac{I_2}{128s^2z^7}
       +\frac{35}{768sz^9}
  \right)\omega^2
  +\left(\frac{\rho_0^{(4)}}{1920}
        +\frac{\rho_0^{(2)}}{768z^4}\right)\omega^4
 \Biggr]\hbar^3+{\cal O}(\hbar^5)\Biggr\}
 \cos\frac{\omega\rho_0}{2}.
\end{aligned}
\label{eq:CDexpres}
\end{align}
Observe that there appear several derivatives
of $\rho_0$ at higher orders. These derivatives
in fact have their origin in the integral \eqref{eq:phi}
and are neatly absorbed if one recasts
the above result into the form \eqref{eq:CD-sincos}.

\section{Conclusions and outlook}\label{sec:conclusion}
In this paper we have computed the higher order corrections of the SFF
in the $\tau$-scaling limit for an arbitrary background $\{t_k\}$
of topological gravity.
We have also shown that these corrections of the SFF can be
obtained by the Fourier transform of 
the CD kernel by including the corrections to the sine kernel formula.
As we can see from Figure \ref{fig:JT-SFF}, $\text{SFF}_{g\geq1}$ approaches a constant at late times
which gives a correction to the value of the plateau.
On the other hand, $\text{SFF}_{g\geq1}$ apparently diverges at $\tau=0$,
but this is just an artifact of the $\tau$-scaling limit, as we argued at the end of
subsection \ref{sec:SFFg}.

There are many interesting open questions.
As far as we know, the corrections of the CD kernel to the sine kernel formula
have not been fully explored in the literature before.
In this paper we have computed these corrections 
of the CD kernel in the double-scaled matrix model of
general topological gravity.
It would be possible to compute the corrections 
of the CD kernel in a random matrix model before taking the double scaling limit.
It would be interesting to 
carry out this computation along the lines of \cite{Brezin:1993qg}.
As emphasized in \cite{Saad:2022kfe,Blommaert:2022lbh},
the $\tau$-scaling limit enables us to 
reproduce the plateau of the SFF by just summing over the 
perturbative genus expansion. In general, the 
perturbative genus expansion is an asymptotic series,
but it becomes a convergent series with a finite radius of convergence
after we take the $\tau$-scaling limit.
What is happening is that, in the $\tau$-scaling limit,
the $(2g)!$ growing part of the
Weil-Petersson volume is suppressed 
and only a non-growing part of the
Weil-Petersson volume survives, and hence
we get a convergent series for the SFF.
In this sense, we throw away most of the contributions of
the moduli space integral by taking the $\tau$-scaling limit.
It is fair to say that we still do not understand the
non-perturbative effects which might contribute to the appearance of the plateau.
In fact, on general grounds we expect that the sum over
$\text{SFF}_g$ in \eqref{eq:SFFg-def} is an asymptotic series. 
It would be interesting to study its resurgence structure 
and understand the non-perturbative effects associated with this asymptotic series
\eqref{eq:SFFg-def}.

\acknowledgments
We are grateful to Douglas Stanford for informing us of the result of
\cite{Saad:2022kfe} prior to its submission to arXiv.
This work was supported
in part by JSPS Grant-in-Aid for Transformative Research Areas (A) 
``Extreme Universe'' 21H05187 and JSPS KAKENHI Grant 19K03856 and 22K03594.

\appendix

\section{Airy kernel}\label{app:airy-kernel}
In this appendix, we consider the CD kernel for the Airy case,
known as the Airy kernel.
As the name suggests, the BA function for the Airy case is given by
the Airy function
\begin{equation}
\begin{aligned}
 \psi(E)=\hbar^{-\frac{1}{6}}\text{Ai}\bigl[-\hbar^{-\frac{2}{3}}(E+t_0)\bigr].
\end{aligned} 
\end{equation}
Plugging this into \eqref{eq:CDkernel}, the Airy kernel is given by
\begin{equation}
\begin{aligned}
 K(E_1,E_2)
&=\frac{\text{Ai}'(-\hbar^{-\frac{2}{3}}E_1)\text{Ai}(-\hbar^{-\frac{2}{3}}E_2)
-\text{Ai}'(-\hbar^{-\frac{2}{3}}E_2)\text{Ai}(-\hbar^{-\frac{2}{3}}E_1)}{E_1-E_2},
\end{aligned} 
\label{eq:airy-kernel}
\end{equation}
where we have set $t_0=0$ after taking the $t_0$-derivative.

To see the corrections to the sine kernel formula,
we first recall that $\text{Ai}(-x)$ and
$\text{Ai}'(-x)$ have the following asymptotic expansion
at large $x>0$,\footnote{See e.g.~\href{https://dlmf.nist.gov/9.7}{https://dlmf.nist.gov/9.7}.}
\begin{equation}
\begin{aligned}
 \text{Ai}(-x)&=\frac{1}{\rt{\pi}x^\qu}\left[
\cos\left(A-\frac{\pi}{4}\right)\sum_{k=0}^\infty(-1)^k\frac{u_{2k}}{A^{2k}}
+\sin\left(A-\frac{\pi}{4}\right)\sum_{k=0}^\infty(-1)^k\frac{u_{2k+1}}{A^{2k+1}}\right],\\
\text{Ai}'(-x)&=\frac{x^\qu}{\rt{\pi}}\left[
\sin\left(A-\frac{\pi}{4}\right)\sum_{k=0}^\infty(-1)^k\frac{v_{2k}}{A^{2k}}
-\cos\left(A-\frac{\pi}{4}\right)\sum_{k=0}^\infty(-1)^k\frac{v_{2k+1}}{A^{2k+1}}\right],
\end{aligned} 
\label{eq:Ai-expand}
\end{equation}
where $A=\frac{2}{3}x^{\frac{3}{2}}$ and
\begin{equation}
\begin{aligned}
 u_k&=\frac{\Ga(k+5/6)\Ga(k+1/6)}{2^{k+1}k!\pi},\\
v_k&=\frac{1+6k}{1-6k}u_k=-\frac{\Ga(k+7/6)\Ga(k-1/6)}{2^{k+1}k!\pi}.
\end{aligned} 
\end{equation}
The first few terms of the expansion \eqref{eq:Ai-expand} read
\begin{equation}
\begin{aligned}
 \text{Ai}(-x)&=\frac{1}{\rt{\pi}x^\qu}\cos\left(A-\frac{\pi}{4}\right)\left[1
-\frac{385}{10368A^2}+\cdots\right]\\
&+\frac{1}{\rt{\pi}x^\qu}\sin\left(A-\frac{\pi}{4}\right)\left[\frac{5}{72A}
-\frac{85085}{2239488A^3}+\cdots\right],\\
\text{Ai}'(-x)&=\frac{x^\qu}{\rt{\pi}}
\sin\left(A-\frac{\pi}{4}\right)\left[1+\frac{455}{10368A^2}+\cdots\right]\\
&-\frac{x^\qu}{\rt{\pi}}
\cos\left(A-\frac{\pi}{4}\right)\left[-\frac{7}{72A}+
\frac{95095}{2239488A^3}+\cdots\right].
\end{aligned}
\label{eq:expand-Ai} 
\end{equation}
The corrections to the sine kernel formula
can be obtained by plugging the above expansion \eqref{eq:expand-Ai}
into \eqref{eq:airy-kernel} with $E_{1,2}=E\pm\hf\hbar\om$.   
By keeping the terms proportional to $\sin(A_1-A_2)$ and $\cos(A_1-A_2)$, we find
\begin{equation}
\begin{aligned}
 K\left(E+\frac{\hbar\om}{2},E-\frac{\hbar\om}{2}\right)
&=\left[\frac{2}{\hbar\om}+\frac{\om}{16z^4}\hbar
+\left(\frac{11 \om^3}{1024 z^8}-\frac{105 \om}{1024 z^{10}}\right)\hbar^3+\cO(\hbar^5)\right]\frac{\sin(A_1-A_2)}{2\pi}\\
&+\left[\frac{\hbar}{16z^5}+\left(\frac{35 \om^2}{768 z^9}-\frac{105}{1024 z^{11}}\right)\hbar^3+\cO(\hbar^5)\right]\frac{\cos(A_1-A_2)}{2\pi},
\end{aligned} 
\label{eq:airy-kernel-hbar}
\end{equation}
where  $z=\rt{E}$ and
\begin{equation}
\begin{aligned}
 A_{1,2}=\frac{2}{3\hbar}E_{1,2}^{\frac{3}{2}}=\frac{2}{3\hbar}
\left(E\pm\hf \hbar\om\right)^{\frac{3}{2}}.
\end{aligned} 
\end{equation}
One can check that \eqref{eq:airy-kernel-hbar} is consistent with our general formula of the CD kernel in \eqref{eq:CD-sincos}.
Note that if we plug \eqref{eq:expand-Ai}
into \eqref{eq:airy-kernel},
there appear terms proportional to $\sin(A_1+A_2)$ and $\cos(A_1+A_2)$
as well. However, they are
highly oscillating in the limit $\hbar\to0$ with fixed $E,\om$, and hence these terms
can be ignored in the computation of the Fourier transform of the CD kernel.

\section{\mathversion{bold}Coefficient $a_2$ of $\text{SFF}_2$
         in \eqref{eq:SFFexp1}}
\label{app:a2}

The coefficient $a_2$ of $\text{SFF}_2$
         in \eqref{eq:SFFexp1} is given by
(for brevity we abbreviate $\zt$ to $z$)
\begin{align}
\begin{aligned}
a_2
&=
\biggl(
 -\frac{e(2)^2}{72e(1)}
 +\frac{1}{18s^3z}
\biggr)\beta^4\\
&\hspace{1em}
+\biggl(
  \frac{e(1)e(2)}{48z^4}
 +\frac{e(4)}{240e(1)}
 -\frac{5e(2)^3}{144e(1)^3}
 +\frac{29I_2}{180s^4z}
 +\frac{e(2)e(3)}{72e(1)^2}
 -\frac{1}{72s^3z^3}
\biggr)\beta^3\\
&\hspace{1em}
+\biggl(
 -\frac{e(2)}{96sz^5}
 -\frac{e(3)}{32z^4}
 +\frac{e(1)e(2)}{16z^6}
 -\frac{e(5)}{160e(1)^2}
 +\frac{e(3)^2}{48e(1)^3}
 +\frac{e(2)^4}{24e(1)^5}
 +\frac{e(2)^2}{32z^4e(1)}\\
&\hspace{2.75em}
 -\frac{3e(1)^3}{64z^8}
 +\frac{7I_2^2}{30s^5z}
 +\frac{29I_3}{360s^4z}
 -\frac{29I_2}{720s^4z^3}
 +\frac{I_2e(2)}{144s^2z^3}
 +\frac{19e(2)e(4)}{480e(1)^3}
 -\frac{31e(2)^2e(3)}{288e(1)^4}\\
&\hspace{2.75em}
 +\frac{1}{96z^5s^3}
\biggr)\beta^2\\
&\hspace{1em}
+\biggl(
  \frac{29I_2I_3}{180s^5z}
 +\frac{e(3)}{96e(1)sz^5}
 -\frac{7I_2e(1)^2}{384s^2z^7}
 +\frac{I_2e(2)}{96s^2z^5}
 -\frac{e(2)^2}{64sz^5e(1)^2}
 -\frac{I_2e(3)}{144s^2z^3e(1)}\\
&\hspace{2.75em}
 +\frac{I_2e(2)^2}{96s^2z^3e(1)^2}
 +\frac{7I_2^3}{36s^6z}
 +\frac{I_4}{48s^4z}
 -\frac{7I_2^2}{120s^5z^3}
 -\frac{29I_3}{1440s^4z^3}
 +\frac{29I_2}{960s^4z^5}
 +\frac{73e(1)^2}{768sz^9}\\
&\hspace{2.75em}
 -\frac{5e(2)}{192sz^7}
 +\frac{e(2)^2}{16z^6e(1)}
 +\frac{e(4)}{64z^4e(1)}
 -\frac{3e(1)^3}{16z^{10}}
 +\frac{13e(2)^5}{24e(1)^7}
 -\frac{e(3)}{16z^6}
 +\frac{e(6)}{320e(1)^3}\\
&\hspace{2.75em}
 -\frac{5}{384s^3z^7}
 -\frac{7e(2)e(3)}{96z^4e(1)^2}
 +\frac{21e(1)e(2)}{128z^8}
 +\frac{e(2)^3}{16z^4e(1)^3}
 +\frac{47e(2)e(3)^2}{144e(1)^5}
 -\frac{73e(2)^3e(3)}{72e(1)^6}\\
&\hspace{2.75em}
 +\frac{181e(2)^2e(4)}{720e(1)^5}
 -\frac{61e(3)e(4)}{960e(1)^4}
 -\frac{113e(2)e(5)}{2880e(1)^4}
\biggr)\beta\\
&\hspace{1em}
 +\frac{25I_2e(1)}{384s^3z^8}
 -\frac{49I_2e(1)^2}{1536s^2z^9}
 -\frac{I_3e(1)}{96s^3z^6}
 -\frac{29I_2I_3}{720s^5z^3}
 -\frac{e(4)}{384sz^5e(1)^2}
 +\frac{5e(3)}{384sz^7e(1)}\\
&\hspace{1em}
 -\frac{e(2)^3}{64sz^5e(1)^4}
 -\frac{5e(2)^2}{256sz^7e(1)^2}
 -\frac{25I_2^2e(1)}{1152s^4z^6}
 +\frac{I_2e(2)}{64s^2z^7}
 +\frac{5I_2^2I_3}{36s^6z}
 +\frac{11I_2I_4}{360s^5z}\\
&\hspace{1em}
 -\frac{I_2e(3)}{192s^2z^5e(1)}
 +\frac{e(2)e(3)}{64sz^5e(1)^3}
 +\frac{I_2e(2)^2}{128s^2z^5e(1)^2}
 +\frac{I_2e(2)^3}{96s^2z^3e(1)^4}
 +\frac{I_2e(4)}{576s^2z^3e(1)^2}\\
&\hspace{1em}
 +\frac{13e(1)e(2)}{64z^{10}}
 +\frac{e(2)^3}{16z^6e(1)^3}
 +\frac{3e(2)^2}{64z^8e(1)}
 +\frac{e(2)^4}{16z^4e(1)^5}
 +\frac{e(3)^2}{64z^4e(1)^3}
 +\frac{e(4)}{64z^6e(1)}\\
&\hspace{1em}
 -\frac{e(5)}{384z^4e(1)^2}
 +\frac{e(2)e(4)}{48z^4e(1)^3}
 -\frac{7e(2)e(3)}{96z^6e(1)^2}
 -\frac{3e(2)^2e(3)}{32z^4e(1)^4}
 -\frac{119e(2)^4e(3)}{48e(1)^8}
 +\frac{7e(2)e(6)}{720e(1)^5}\\
&\hspace{1em}
 +\frac{29e(3)e(5)}{1440e(1)^5}
 +\frac{e(4)^2}{80e(1)^5}
 -\frac{e(7)}{1920e(1)^4}
 +\frac{7e(2)^6}{6e(1)^9}
 -\frac{7e(3)^3}{96e(1)^6}
 -\frac{9e(2)}{128sz^9}
 -\frac{53e(1)}{512s^2z^{10}}\\
&\hspace{1em}
 +\frac{219e(1)^2}{1024sz^{11}}
 -\frac{29I_2}{768s^4z^7}
 +\frac{29I_3}{1920s^4z^5}
 -\frac{I_4}{192s^4z^3}
 +\frac{7I_2^2}{160s^5z^5}
 -\frac{7I_2^3}{144s^6z^3}
 +\frac{I_5}{288s^4z}\\
&\hspace{1em}
 +\frac{29I_3^2}{1440s^5z}
 +\frac{7I_2^4}{72s^7z}
 -\frac{23e(2)e(3)e(4)}{72e(1)^6}
 +\frac{7e(2)^3e(4)}{12e(1)^7}
 +\frac{19e(2)^2e(3)^2}{16e(1)^7}
 -\frac{3e(2)^2e(5)}{32e(1)^6}\\
&\hspace{1em}
 -\frac{15e(1)^3}{64z^{12}}
 -\frac{15e(3)}{256z^8}
 +\frac{35}{1536s^3z^9}
 -\frac{I_2e(2)e(3)}{96s^2z^3e(1)^3}.
\end{aligned}
\label{eq:a2}
\end{align}
One can check that the $e(j)$-dependent part of \eqref{eq:a2} reproduces 
$h_{2,n}$ in \eqref{eq:h2n}; the $e(j)$-independent part of \eqref{eq:a2}
becomes $d_2(z_\tau,\bt)$ defined in \eqref{eq:ZErf-cg}. 


\bibliography{paper}
\bibliographystyle{utphys}

\end{document}